\documentclass{article} % For LaTeX2e
\usepackage{iclr2026_conference,times}

\usepackage{amsmath,amsfonts,bm}

\def\eqref#1{equation~\ref{#1}}
\def\1{\bm{1}}

\DeclareMathAlphabet{\mathsfit}{\encodingdefault}{\sfdefault}{m}{sl}
\SetMathAlphabet{\mathsfit}{bold}{\encodingdefault}{\sfdefault}{bx}{n}

\usepackage{hyperref}
\usepackage{url}

\usepackage{amsthm}
\usepackage{graphicx}
\usepackage{booktabs}
\usepackage{enumitem}
\usepackage{amssymb}
\usepackage{caption}
\usepackage{subcaption}
\usepackage{wrapfig}
\usepackage[english]{babel}
\usepackage{mathabx}
\usepackage{tikz}
\usetikzlibrary{decorations.pathmorphing, arrows.meta}

\usepackage{multirow}
\usepackage{xspace}
\newcommand{\postgres}[0]{\textsc{PostgreSQL}\xspace}
\newcommand{\true}[0]{\textsc{Oracle}\xspace}
\newcommand{\mscn}[0]{\textsc{MSCN}\xspace}
\newcommand{\deepdb}[0]{\textsc{DeepDB}\xspace}
\newcommand{\factorjoin}[0]{\textsc{FactorJoin}\xspace}
\newcommand{\price}[0]{\textsc{PRICE}\xspace}
\newcommand{\ours}[0]{\textsc{LiteCard}\xspace}
\newcommand{\sparagraph}[1]{\vspace{2mm} \noindent \textbf{#1} } 

\newtheorem{theorem}{Theorem}

\newtheorem{definition}{Definition}

\title{
Is it Bigger than a Breadbox: %learning
Efficient Cardinality Estimation for Real World Workloads
}

\author{
Zixuan Yi\thanks{Major Contributions.  \hspace{0.5cm}  $^\dagger$ Work performed at Google, as a Student Researcher. \hspace{0.5cm} $^\ddagger$ Now at ETH.}\ \ $^\dagger_{(P)}$ , Sami Abu-el-Haija$^{*}_{(G)}$ \\
\textbf{Yawen Wang$_{(G)}$, Teja Vemparala$_{(G)}$, Yannis \ Chronis$^\ddagger_{(G)}$, Yu Gan$_{(G)}$, Michael Burrows$_{(G)}$} \\
\textbf{Carsten Binnig$_{(D)}$, Bryan Perozzi$_{(G)}$, Ryan Marcus$_{(P)}$, Fatma \"Ozcan$_{(G)}$} \vspace{0.3cm} \\ 
(P): University of Pennsylvania; \hspace{2cm}
(G): Google; \hspace{2cm}
(D): TU Darmstadt
}

\iclrfinalcopy % Uncomment for camera-ready version, but NOT for submission.
\begin{document}

\maketitle

\begin{abstract}
DB engines produce efficient query execution plans by relying on cost models. Practical implementations estimate cardinality of queries using heuristics, with magic numbers tuned to improve average performance on benchmarks. Empirically, estimation error significantly grows with query complexity.
Alternatively,
learning-based estimators offer improved accuracy, but add operational complexity preventing their adoption in-practice.
Recognizing that query workloads contain highly repetitive subquery patterns, 
we learn many simple regressors online, each localized to a pattern. The regressor corresponding to a pattern can be randomly-accessed using hash of graph structure of the subquery.
Our method has negligible overhead and competes with SoTA learning-based approaches on error metrics. Further, amending PostgreSQL with our method achieves notable accuracy and runtime improvements over traditional methods and drastically reduces operational costs compared to other learned cardinality estimators, thereby offering the most practical and efficient solution on the Pareto frontier.
Concretely, simulating JOB-lite workload on IMDb speeds-up execution by 7.5 minutes ($>$30\%) while incurring only 37 seconds overhead for online learning.
\end{abstract}

\section{Introduction}
% Why does the world care.
The majority of computer applications of any significant utility use relational databases. Performance optimization of query execution has therefore been studied for decades, \textit{e.g.}, \citet{astrahan76, selinger79, graefe1987exodus, IoannidisNSS97, trummer2015multi}.
\textbf{Cardinality Estimation} -- the task of predicting the record-count of (sub-)queries -- is essential for query plan optimization \citep{leis2015good, Marcus:2021:Bao, lee2023analyzing}.

The popular database engine, PostgreSQL,
estimates cardinalities using
per-column histograms \citep{postgres_estimator},
na\"ively assuming that columns are uncorrelated.
Advantages of this heuristic include its speed-of-calculation, which allows it to be invoked numerous times for multi-join queries.
%, as well as its applicability to any query.
However, this estimation exhibits large errors when independence assumptions are violated, \textit{e.g.}, when joining records from multiple tables, unnecessarily slowing-down query execution  by possibly orders-of-magnitudes \citep{moerkotte2010preventing}.

A variety of deep-learning methods propose to capture intricate data distributions, either directly by sampling records \citep[\textit{e.g.}, ][]{deepdb, factorjoin}, or indirectly by posing \textit{cardinality estimation} as a supervised learning task \citep[\textit{e.g.}, ][]{mscn, cardbench}. While these models can discover correlations across columns and produce better cardinality estimates than heuristic algorithms, their overheads prevents their adoption in practice \citep{wang20areweready}.

%\textcolor{red}{This paragraph is copied from previous Related Work section. TODO: Adopt into here.} The limitations of existing approaches in terms of practical deployment overheads – including initial training, query optimization latency, and model update costs – motivate the need for a new framework. Specifically, a practical learned cardinality estimator should ideally offer low-to-zero overhead for a new database, minimal optimization time per query, and fast, continuous updatability to adapt to evolving workloads and data, while providing performance comparable to or better than the state-of-the-art learned approaches. This is the gap our work aims to fill.

In this paper, we strive to design a cardinality estimator that: (i) can run from cold-start, requiring no upfront training; (ii) can adapt to changes in workloads or data shifts; and (iii) has negligible update and inference time.
We propose such an estimator.
Rather than a monolithic neural network that processes all queries, we employ many small models,
each specializes to one sub-query pattern.
The query pattern is identified from the \textit{structure} of the graph corresponding to the query, while excluding some node features, \textit{e.g.}, constant values, table names and/or column names.
Our proposed method fits within a general a class of learning methods known as \textit{locally-weighted models}.
Prediction on any data point requires fitting a new model on training examples that are near the data point.
These methods define a (similarity) \textit{Kernel} function, that generally operates on pairs of \textbf{numeric} feature vectors. However, our kernels integrate \textbf{both} the \textbf{graph structure and numeric} data.
%
%\textcolor{red}{incomplete.}

%The paper continues as follows. We review background material in \S\ref{sec:background}, while incrementally introducing our notation. Our main contribution is in \S\ref{sec:ourmethod_graphlocal} and \S\ref{sec:ourmethod_implementation}. We compare the estimation error of our method versus classical and modern methods in \S\ref{sec:}. Furthermore, we show the runtime savings of our method when installed within PostgreSQL in \S\ref{sec:}, while comparing with alternatives.

\section{Background}
\label{sec:background}
\subsection{Graph Representation of (Sub)queries and Query Plan Optimization}

Database engines rely on \textit{cost models}
to create efficient \textit{query execution plan} for responding to a query.
The plan is a tree: leaf-nodes read data records, generally from table columns, and as the data traverses down the tree, records get merged (per joined columns) and filtered (per predicates), finally producing one record stream at the root, \textit{i.e.}, the response to the query.
There can be many valid plans for a query. However, some plans are favored, requiring fewer resources and executing faster.
While searching for an optimal plan, the cost model must 
estimate the cardinality of candidate sub-queries (nodes) before they get selected into the query plan (tree). The cardinality is the number of records output by the subquery (emitted by the node, down the tree).
Consider the simple SQL:
%\footnote{For broader readibility, we depict our data structures on SQL, mirroring our Python implementation. Please refer to Appendix for PostgreSQL QueryPlans (``\texttt{RelInfo}''), mirroring our C++ implementation in Postgres.}:
\begin{equation}
\texttt{\textbf{SELECT} \dots  \textbf{ FROM} movies \textbf{WHERE} stars>3 and year IN (2024,2025)}
\label{eq:sample_sql}
\end{equation}
The statement queries movies produced in the last 2 years, rated above 3-stars.
Let us assume that both columns, \texttt{stars} and \texttt{year}, are individually indexed but are not co-indexed.
Then, the Query Plan Optimizer estimates the cardinality of two constituent sub-queries:
\begin{equation*}
\fbox{\texttt{\textbf{SELECT}\dots\textbf{WHERE} stars > 3}}
\ \ \ \textrm{ and } \ \ \ 
\fbox{\texttt{\textbf{SELECT}\dots\textbf{WHERE} year IN (2024, 2025)}}
\end{equation*}
The optimizer uses cardinality estimates to determine the \textit{join type}.
For instance, if the second subquery has a low cardinality estimate, then it could be executed earlier, and its (primary-key, record) outputs can be stored in-memory before the first subquery executes. However, if both subqueries have large cardinalities, then they can be separately executed, sorted by primary key, then intersected in a streaming-fashion. These are respectively named \textit{broadcast join} and \textit{merge join}.
% Complexity increases with multi-table-joins.
%
Cardinality estimation also  determines \textit{join orders}. For instance, when joining 3 tables (A$\bowtie$B$\bowtie$C), the optimizer must choose which two tables merge first  ((A$\bowtie$B$)\bowtie$C) or  (A$\bowtie($B$\bowtie$C)).
The number of join orderings can be  exponential in the number of tables.
While searching for the optimal plan, the optimizer repeatedly invokes the cardinality estimator, \textit{e.g.}, up to thousands of times for complex queries.

% \subsection{Query Optimization in Database Engines}
% Summarize Query (plot)
% Summarize QueryPlan (plot)
% Search tree.

\begin{wrapfigure}{r}{0.25\textwidth}
\vspace{-0.5cm}
    \centering
    \includegraphics[width=0.9\linewidth]{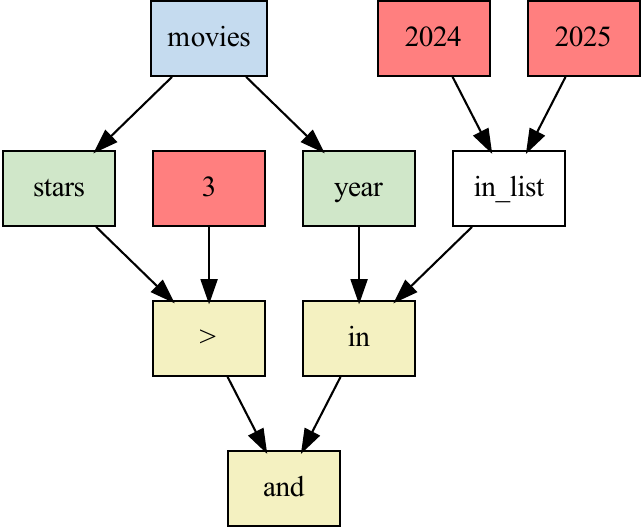}
    \caption{DAG corresponding to SQL in Eq.~\ref{eq:sample_sql}}
    \label{fig:annotated_sql}
\end{wrapfigure}
\paragraph{Graph Representation of (sub)queries.}
Queries are generally represented as trees in database engines \citep{rulerewrite, encyclopediadb, RamakrishnanGehrke03}, and we convert them to directed acyclic graphs (DAGs) similar to \citet{cardbench}. Details are in appendices \ref{appendix:sql_to_dag}\&\ref{appendix:postgres_to_dag}.
Figure~\ref{fig:annotated_sql} depicts such a DAG. There are different node types, each type has its own feature sets and is depicted with a different color.
Let $\mathcal{T}$ denote the universe\footnote{
Entries listed in $\mathcal{T}$ and $\mathcal{A}$ are not exhaustive. DB engineers may keep additional information helpful for modeling, \textit{e.g.}, number of unique values per column, min- and max-column values, histograms, bloom-filters...
} of node types that can appear in the (sub)query graph. In our application,
\begin{equation}
    \mathcal{T} = \{ \underbrace{\textit{table}, \textit{alias}, \textit{column}, \textit{literal},  \textit{op}, \textit{function}}_{\textrm{for graphs extracted from SQL or PostgreSQL's RelInfo}} \ , \underbrace{\textit{join}, \textit{scan}, ..}_{\textrm{for PostgreSQL's}} \}
\end{equation}
For algorithmic correctness, all sets $\{.\}$ are ordered. Let $\mathcal{A}$ be set of pairs (type, attribute name):
\begin{equation}
    \mathcal{A} = \{ (\textit{table}, \textit{name}), (column, name), (column, type), (literal, value), (op, code), \dots  \}
\end{equation}

% In our work, we develop a more-general SQL-to-DAG parser than the one of
% \citet{cardbench}. Further, we provide instructions how to convert tree-like  \texttt{RelInfo} of Postgres into DAG in Appendix \ref{sec:appendix_graphs}. 

% \begin{figure}
%     \centering
%     \includegraphics[width=\linewidth]{figs/annotated_sql.pdf}
%     \caption{Caption}
%     \label{fig:annotated_sql}
% \end{figure}

%\subsection{Subquery Pattern Repetition}
% Notwithstanding workloads with diverse queries, sub-query patterns tend to repeat.
% \textcolor{red}{Show some examples.}

\subsection{Localized Models}
\label{sec:localmodels}
\textbf{Local models} infer on a data point $\mathbf{x}$ by considering  \textbf{nearby} points. 
Proximity between points $\mathbf{x}$ and $\mathbf{z}$ is measured by kernel function $K(\mathbf{x}, \mathbf{z}) \ge 0$.
A notable choice is the Gaussian kernel with
\begin{equation}
K_\sigma(\mathbf{x}, \mathbf{z}) = \exp\left(-\frac{||\mathbf{x}-\mathbf{z}||^2}{\sigma^2}\right) \in [0, 1]
\label{eq:k_rbf}
\end{equation}

where hyperparameter $\sigma > 0$ is known as the \textit{kernel width} or \textit{variance}.
This kernel frequently appears. We utilize it in two ways. First, in \textit{locally-weighted linear regression} \citep{Cleveland1979},
%where inference for test point $\mathbf{x}$ requires fitting a weighted regression model utilizing previous observation points
%$\{\mathbf{z}^{(i)}\}_{i \in \textrm{history}}$ with weights $\{K_\sigma(\mathbf{x}, \mathbf{z}^{(i)})\}_i$.
Second,
in one-shot prediction \citep{hechenbichler2004weighted}.
%, where inference is a linear combination of observed target values $\{y^{(i)}\}_i$ with weights $\{K_\sigma(\mathbf{x}, \mathbf{z}^{(i)})\}_i$.

%\subsection{Model Per-Template}

% \textbf{Per Template Models.} Since queries and subqueries tend to repeat in patterns \citep{redshift}, we are inspired to split a (sub)query graph into a template and parameters, e.g., similar to \citet{blackbox-cardinality}, 
%but differently operating in the graph representation of the (sub)query, additionally offering invariance to permutation ordering (\textit{e.g.}, of conjunctions) and aliases (\textit{e.g.}, of table names).

\definecolor{h10561837651767186983}{HTML}{DAA520}
\definecolor{h11750077659592055130}{HTML}{800000}
\definecolor{h13449585261118054869}{HTML}{0000CD}
\definecolor{h15763554885717590606}{HTML}{EE82EE}
\definecolor{h17502845754571064274}{HTML}{D2691E}
\definecolor{h5282864164141950681}{HTML}{00CED1}

\begin{figure}[t]
    \hspace{0.25cm}
    \begin{minipage}{0.38\linewidth}
    \includegraphics[width=\linewidth]{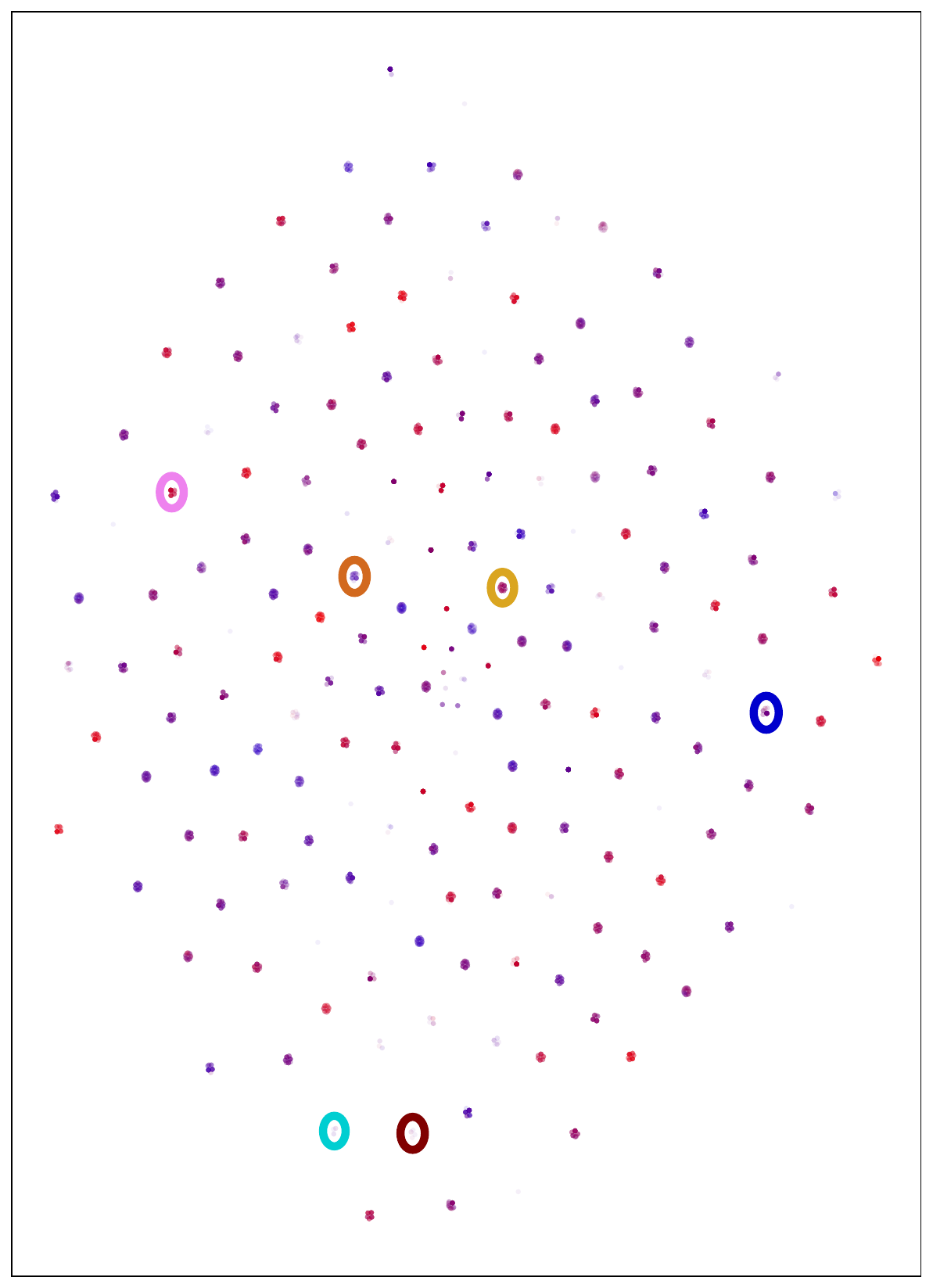}   
    \end{minipage}~
    \begin{minipage}{0.52\linewidth}
        \fcolorbox{h10561837651767186983}{white}{\begin{minipage}{0.49\linewidth}\centering
        \includegraphics[ height=0.9cm]{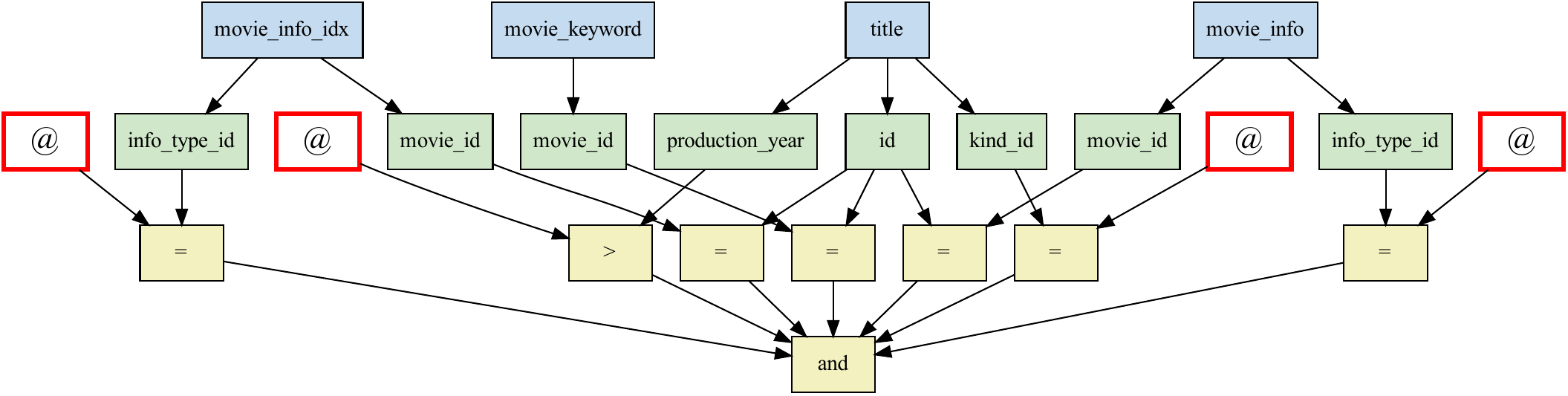}
        \newline
        \includegraphics[width=\linewidth, height=1.2cm]{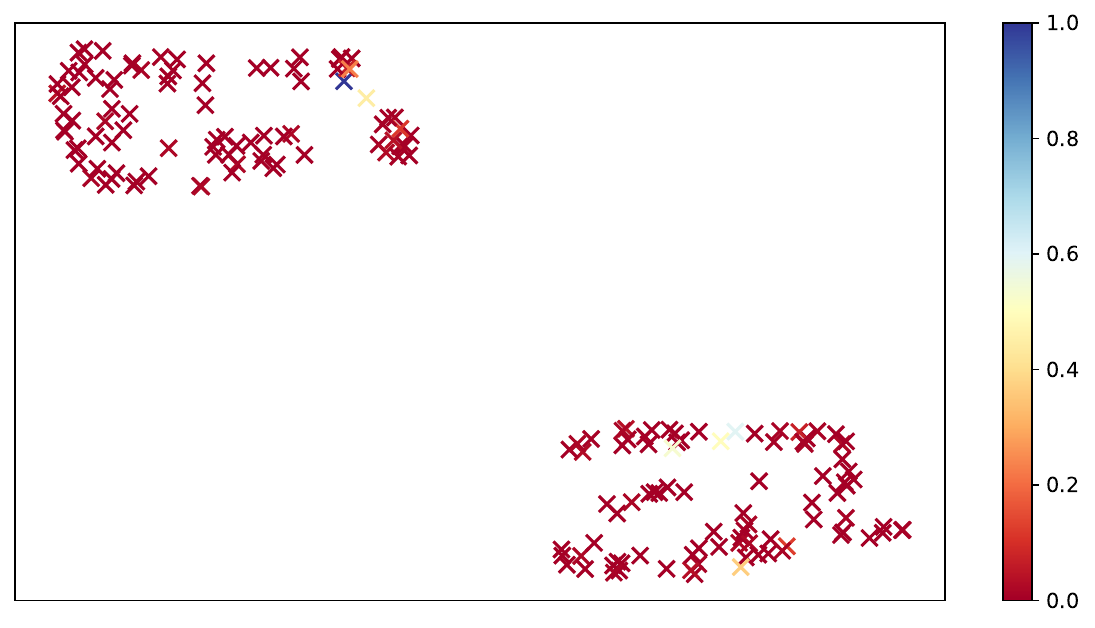}    
        \end{minipage}}~
        \fcolorbox{h11750077659592055130}{white}{\begin{minipage}{0.49\linewidth}\centering
        \includegraphics[ height=0.9cm]{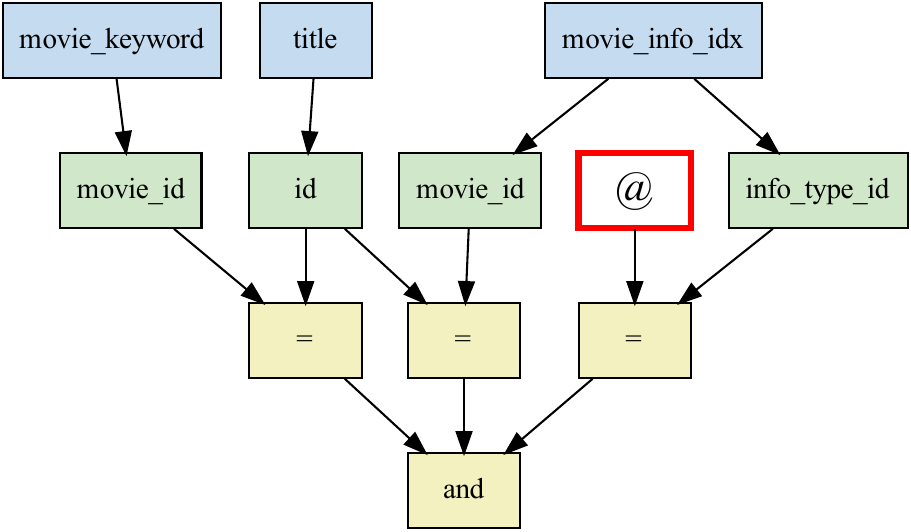}
        \newline
        \includegraphics[width=\linewidth, height=1.2cm]{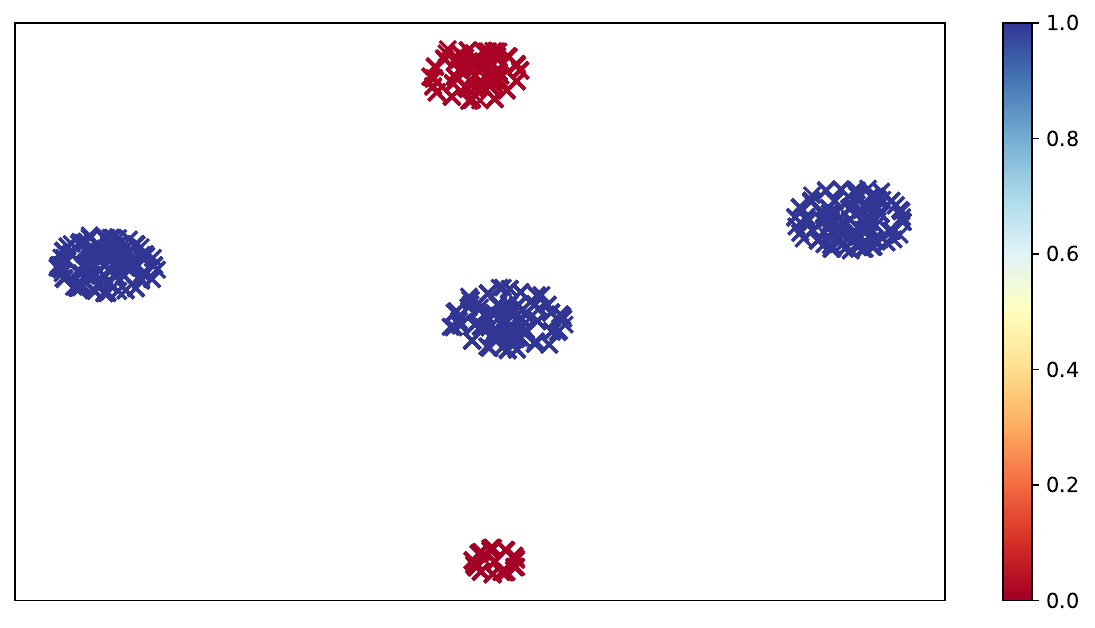}    
        \end{minipage}}
        %%%%%%%%%%%%%
        \newline
        \fcolorbox{h13449585261118054869}{white}{\begin{minipage}{0.49\linewidth}\centering
        \includegraphics[ height=0.9cm]{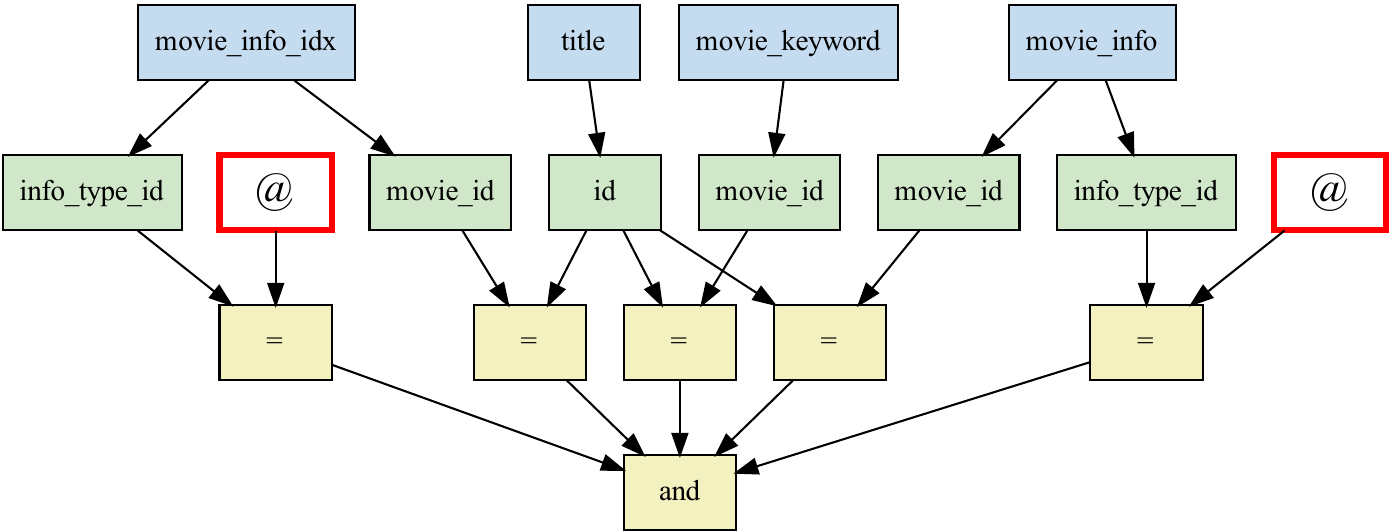}
        \newline
        \includegraphics[width=\linewidth, height=1.2cm]{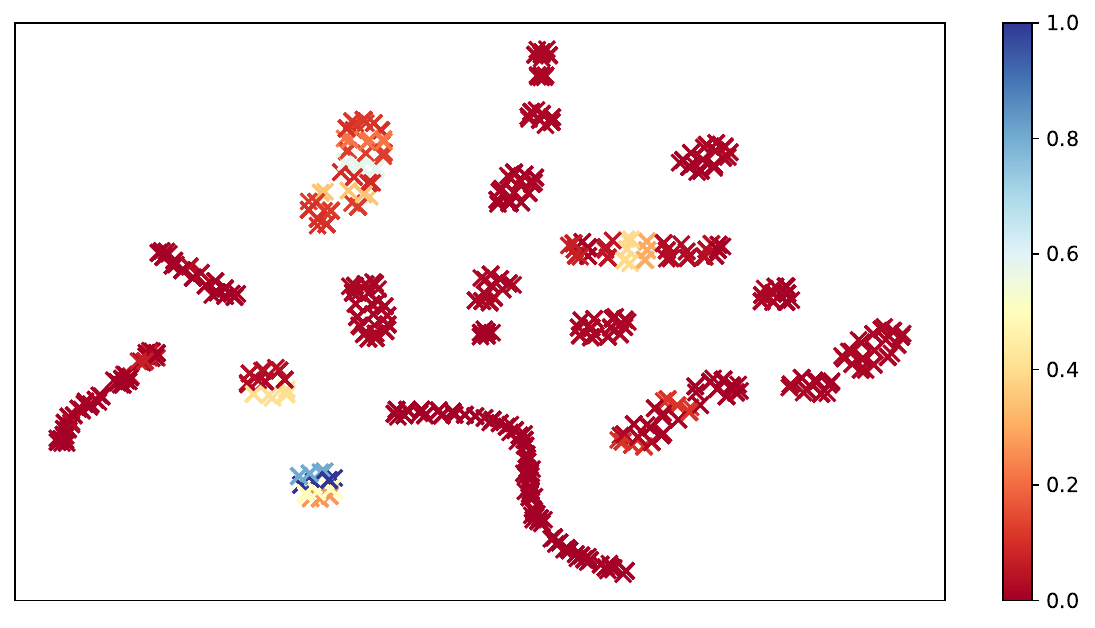}    
        \end{minipage}}~
        \fcolorbox{h15763554885717590606}{white}{\begin{minipage}{0.49\linewidth}\centering
        \includegraphics[ height=0.8cm]{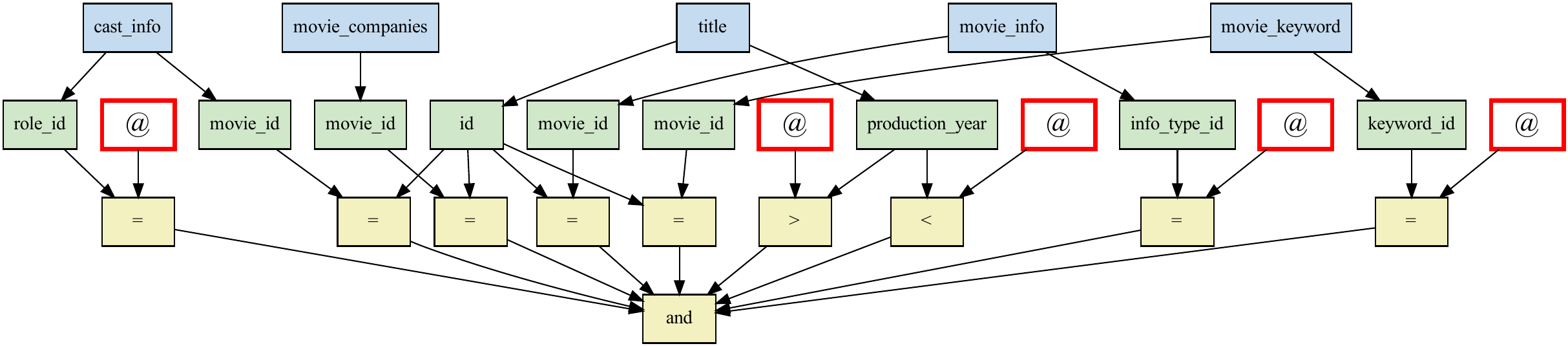}
        \newline
        \includegraphics[width=\linewidth, height=1.2cm]{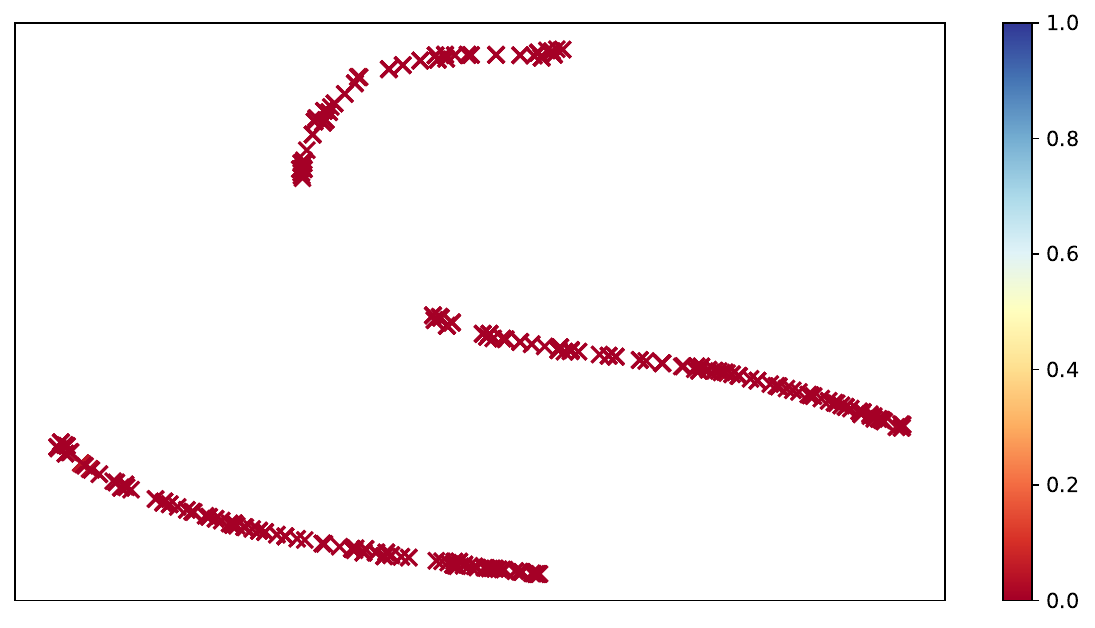}    
        \end{minipage}}
        %%%%%%%%%%%%%
        \newline
        \fcolorbox{h17502845754571064274}{white}{\begin{minipage}{0.49\linewidth}\centering
        \includegraphics[ height=0.9cm]{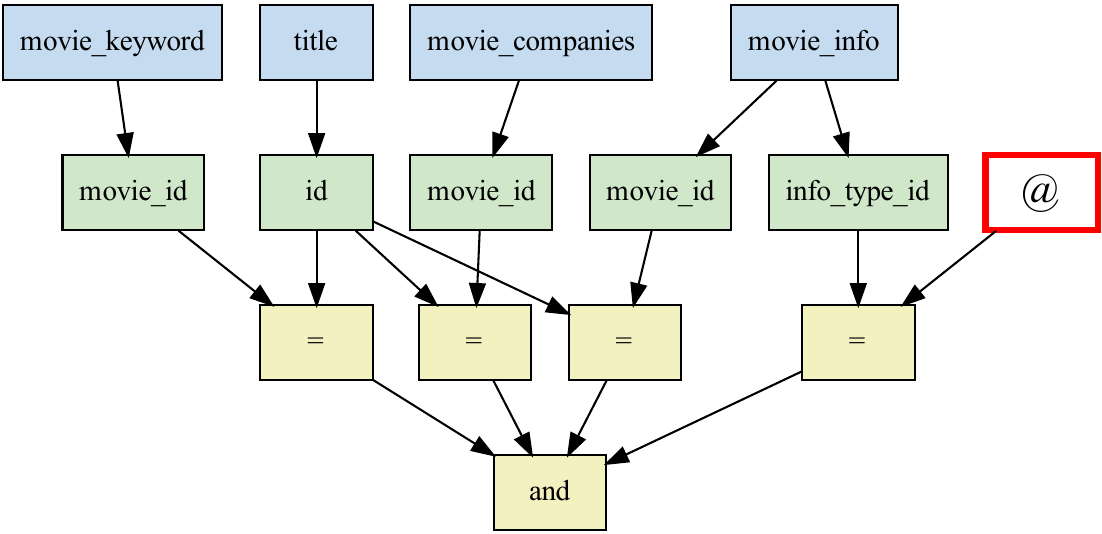}
        \newline
        \includegraphics[width=\linewidth, height=1.2cm]{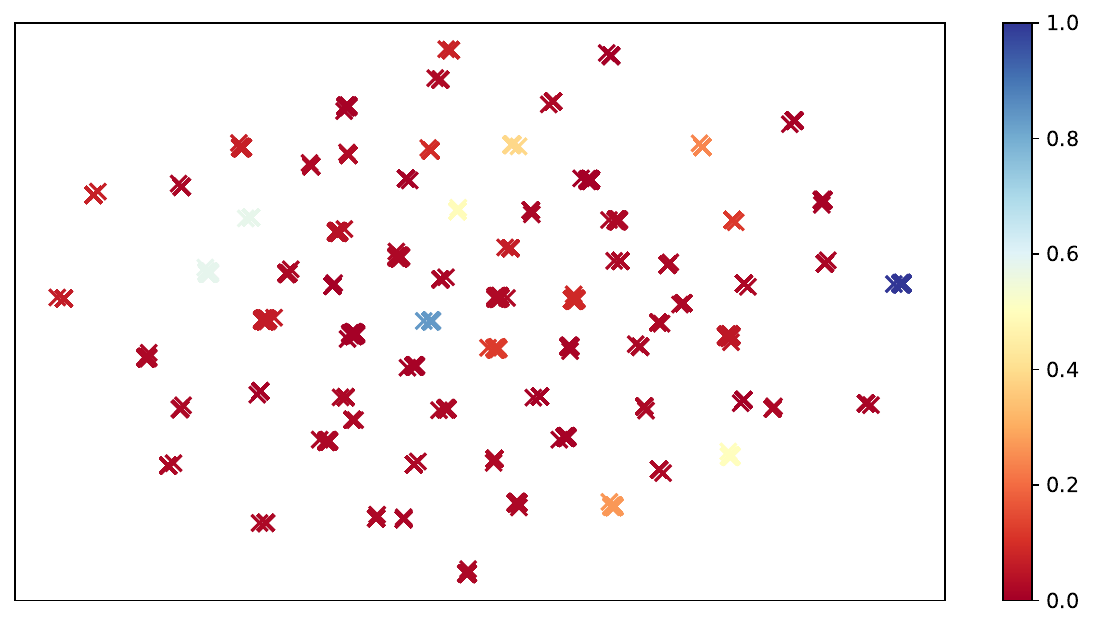}    
        \end{minipage}}~
        \fcolorbox{h5282864164141950681}{white}{\begin{minipage}{0.49\linewidth}\centering
        \includegraphics[ height=0.9cm]{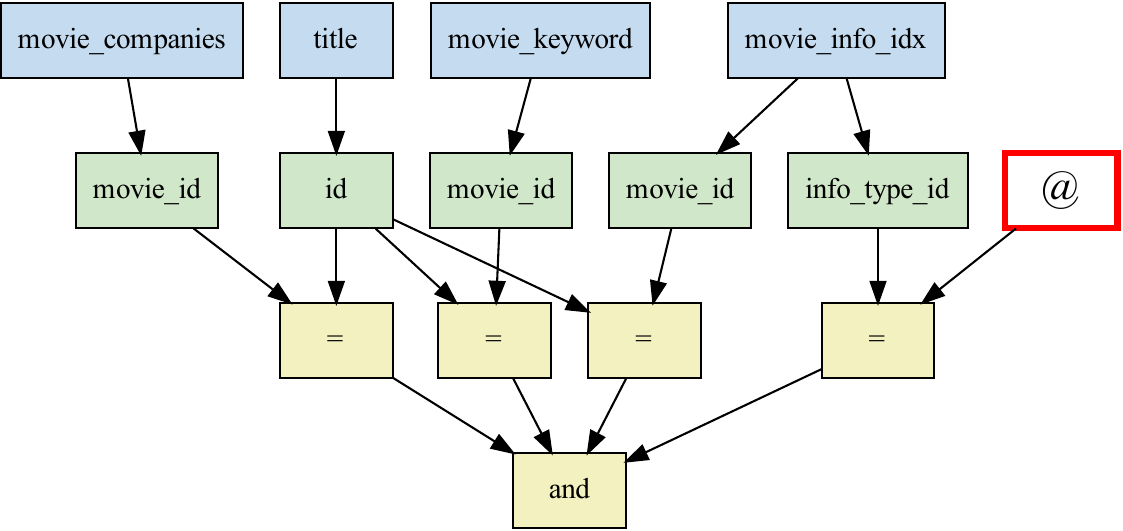}
        \newline
        \includegraphics[width=\linewidth, height=1.2cm]{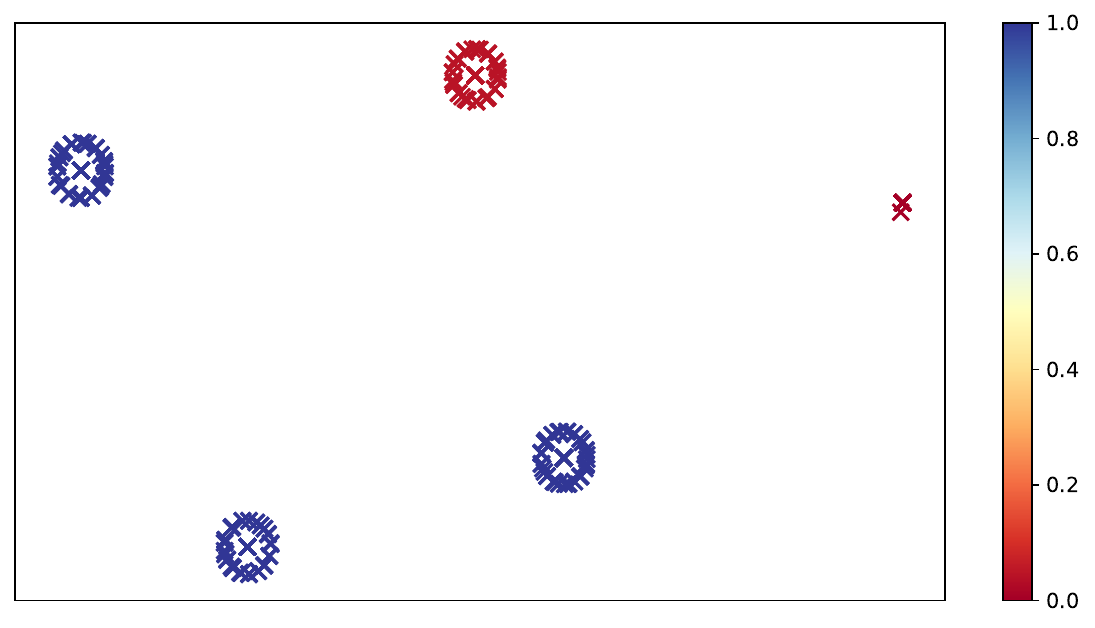}    
        \end{minipage}}
        %%%%%%%%%%%%%
        \newline

    \end{minipage}
    \caption{t-SNE visualizations of IMDB 5K workload. \textbf{(Left)} Every subquery is a point (with 5\% opacity). Due to $K^\mathcal{H}_\mathcal{F}(G, G') = \mathbf{1}_{[h^\mathcal{H}(G) = h^\mathcal{H}(G')]} \times . $, subquery DAGs that are isomorphic (per $\mathcal{H})$ are cleanly clustered, painting a darker region. The point color represents cardinality of the query (from red to blue). We choose 6 clusters (by stratified sampling) and circle them with colors.
    \textbf{(Right)} we recompute t-SNE \textbf{within} each colored cluster. The original dimension of every right plot equals the number of \fcolorbox{red}{white}{@} nodes in the graph above it, which renders the subquery pattern graph.
    Finally, points are colored using their ground-truth (normalized) cardinalities.
    }
    \label{fig:placeholder}
\end{figure}

\section{Graph-Local Learning}
\label{sec:ourmethod_graphlocal}

We first present our final model, top-to-bottom, and the remainder of the section provides details.

Let $(\mathcal{G}' , y') \in \mathcal{D}$ denote history of previously-seen (sub)query DAGs, each associated with its cardinality. % $y \in \mathbb{Z}_+$.
History $\mathcal{D}$ starts empty and populates while queries are executing.
Fig.~\ref{fig:annotated_sql} captures three such DAGs, each rooted at a yellow node.

Inspired by \S\ref{sec:localmodels}, given a test (sub)query graph $\mathcal{G}$, we estimate its cardinality by inference:
\begin{equation}
    g_{\theta}(\mathcal{G}) \ \ \ \ \textrm{  where  }  \ \ \ \
    \theta = \mathop{\arg\min}_{\theta'} \sum_{(\mathcal{G}' , y') \in \mathcal{D}}  K^{\mathcal{H}}_{\mathcal{F}} (\mathcal{G}, \mathcal{G}') \times  (g_{\theta'}(\mathcal{G'}) - y')^2,
\label{eq:localmodel}
\end{equation}
The hyperparameters \textbf{pattern features} $\mathcal{H} \subset \mathcal{A}$ and \textbf{learning features} $\mathcal{F} \subset \mathcal{A}$ are explained in \S\ref{sec:hash_and_feat}.
Kernel $K_\mathcal{F}^{\mathcal{H}}(., .) \ge 0$ outputs large value if its inputs are similar, both feature- and structure-wise, as:
\begin{equation}
    K^{\mathcal{H}}_{\mathcal{F}} \left(\mathcal{G}, \mathcal{G}'\right) =
    \underbrace{\mathbf{1}_{\left[h^\mathcal{H}(\mathcal{G})= h^\mathcal{H}(\mathcal{G'})\right]}}_{G\&G' \textrm{ are isomorphic}}
    \times
    \underbrace{K_\sigma\left(\mathbf{x}^{\mathcal{H}}_{\mathcal{F}} (\mathcal{G}), \mathbf{x}^{\mathcal{H}}_{\mathcal{F}} (\mathcal{G}') \right)}_{\textrm{their features are nearby }}
    \label{eq:kernel}
\end{equation}
where  $K_\sigma$ is defined in Eq.~\ref{eq:k_rbf}  and $\mathbf{x}^{\mathcal{H}}_{\mathcal{F}} (\mathcal{G})$ denotes a feature vector containing features listed in $\mathcal{F}$ from $\mathcal{G}$'s nodes,  respecting canonical node-ordering established by
$\mathcal{H}$. Indicator function $\mathbf{1}_{\left[h^\mathcal{H}(\mathcal{G})= h^\mathcal{H}(\mathcal{G'})\right]}$ evaluates to 1 when $\mathcal{G}$ and $\mathcal{G}'$ are isomorphic when considering features $\mathcal{H}$, and to 0 otherwise.

% and $K^{\mathcal{H}}_{\mathcal{F}}(., .) \ge 0$ compares dense vectors extracted from nodes per feature set $\mathcal{F}$,
% while using $\mathcal{H}$ for canonical node ordering.
The model $g_{\theta}$ is fit locally around $\mathcal{G}$. We restrict ourselves to simple models that can quickly train with negligible overheads. We experiment with Locally-weighted Linear Regression $g^\textrm{LR}_{\theta}$, in addition to Gradient-boosted Decision Forests $g^\textrm{DF}$ (we use implementation of \citep{ydf}).
For conciseness, we ignore the regularization terms from Eq.~\ref{eq:localmodel}, such as $\ell_2$ regularization for Linear Regression, or height-restriction for Decision Forests.
Furthermore, we experiment with one-shot predictors following \citet{hechenbichler2004weighted}, with:
\begin{equation}
    g^\textrm{RBF}(\mathcal{G}) =  \frac{1}{Z} \sum_{(\mathcal{G}' , y')\in \mathcal{D}}  K^{\mathcal{H}}_{\mathcal{F}} (\mathcal{G}, \mathcal{G}') \times   y'
\ \ \textrm{with}   \ \
Z = \sum_{(\mathcal{G}' , y') \in \mathcal{D}}  K^{\mathcal{H}}_{\mathcal{F}} (\mathcal{G}, \mathcal{G}')
\label{eq:oneshot}
\end{equation}

\textbf{System Integration.} We implement functions $g(.)$ and $K^{\mathcal{H}}_{\mathcal{F}}(., .)$  in open-source PostgreSQL (details are in \S \ref{appendix:postgres_to_dag}).
The Query Planner invokes
them 
while searching for the optimal plan.
Once the plan is finalized then executed, cardinalities of all subgraphs (yellow nodes of Fig.~\ref{fig:annotated_sql}) are recorded in $\mathcal{D}$.
% As $\mathcal{D}$ grows, the implementation (\S\ref{sec:ourmethod_implementation}) starts trusting estimates of 
% $g(\mathcal{G})$ when subquery $\mathcal{G}$ is \textit{familiar}.
% In particular, when size of 
% $
% \{  (\mathcal{G}', y')  \in \mathcal{D} \  \mid  \ K^{\mathcal{H}}(\mathcal{G}, \mathcal{G}') = 1 \}
% $ is larger than a threshold.

\subsection{Definitions}

% Our modular design allows ablations of which features may improve cardinality estimation and hence reduce benchmark run-times, for instance, to guide decisions of DB engineers whether or not to include the calculation of certain features such as column-wise statistics.

Let $\{0, 1\}^k$ be a string with $k$ bits and let $\{0, 1\}^*$ be a string with arbitrary length. We denote a (cryptographic) 256-bit hash $\$ : \{0, 1\}^* \rightarrow \{0, 1\}^{256}$.
Let $\mathcal{G}=(\mathcal{V}, \mathcal{E}, \mathcal{X})$ represent a query graph (depicted in Fig.~\ref{fig:annotated_sql}), with node set $\mathcal{V}=\{1, 2, \dots, n\}$ where $n$ denotes number of nodes ($n$=10 in Fig.~\ref{fig:annotated_sql}). Edge set $\mathcal{E} \subset \mathcal{V} \times \mathcal{V}$ contains directed edges ($|\mathcal{E}|$=10 in Fig.~\ref{fig:annotated_sql}) that must necessarily induce a \textit{directed acyclic graph} (DAG). Reverse edge set $\mathcal{E}^\top = \{(v, u)\}_{(u, v) \in \mathcal{E}}$.
The feature set $\mathcal{X} \in ( \mathcal{A} \mapsto \{0, 1\}^*)^n$ stores multiple features per node.
%e.g., name and type for \texttt{Column}.
$\mathcal{X}_j[(t, a)]$ denotes accessing string-valued attribute $(t, a) \in \mathcal{A}$ for node $j \in \mathcal{V}$. Suppose
$(t, a) = (\textit{table}, \textit{name})$ and $j$ corresponds to the index of blue node of Fig.~\ref{fig:annotated_sql}, then $\mathcal{X}_j[(t, a)] =$ ``\texttt{movies}''. If node $j$ does not have attribute $(t, j)$ then $\mathcal{X}_j[(t, a)]$ defaults to null (or empty-string).
Let $\tau_j \in \mathcal{T}$ denote the type of node $j \in \mathcal{V}$.

\subsection{Canonical Ordering, Hashing and Feature Extraction}
\label{sec:hash_and_feat}
\textbf{Canonical Ordering and Pattern Hashing.}
$\mathcal{H} \subset \mathcal{A}$ can effectively
partition incoming queries online.
We first assemble an array of strings $\mathbf{H} \in \{0, 1 \}^{n \times 256}$ with row $j \in \mathcal{V}$ initialized as:
\begin{equation}
    \mathbf{H}_j^\mathcal{H} := \$( \oplus\{\mathcal{X}_j[(t, a)] \mid \tau_j = t \}_{(t, a) \in \mathcal{H}} )
\label{eq:H_j}
\end{equation}
% \begin{equation}
%     \mathbf{H}^\mathcal{H} := \left\{ \textrm{\Large \$} \left( \bigoplus_{(t, a) \in H} \mathcal{X}_j[(t, a)]  \right) \right \}_{j \in \mathcal{V}} \in \{0, 1 \}^{n \times 256},
% \end{equation}
where $\oplus\{.\}$ denotes string-concatenation of elements in ordered set $\{.\}$.
The hash value $\mathbf{H}_j^\mathcal{H} \in \{0, 1\}^{256}$  at this initialization $\approx$uniquely\footnote{
If we assume \$ is a uniform cryptographic hash function, then expected collision rate $\approx\frac{\textrm{UniqPatterns}}{2^{256}}$ .
%even $^{(**)}$ if  function \$(.) consumes its output,  (Eq.~\ref{eq:H_j}) then topologically-ordered updates (Eq.~\ref{eq:H_j_update}).
} identifies node $j$'s feature values, while restricting to pattern features $\mathcal{H}$.
Then, we update the entries:
\begin{align}
    \mathbf{H}_j^\mathcal{H} &:= \$ \left(
    \mathbf{H}_j^\mathcal{H} \oplus \texttt{sort}( \{ \mathbf{H}_k^\mathcal{H} \mid (k, j) \in \mathcal{E} \})   \right)  \ \ \forall j \in \texttt{TopologicalOrder}(\mathcal{E}), \textrm { then }, \label{eq:H_j_update} \\
    \mathbf{H}_j^\mathcal{H} &:= \$ \left(
    \mathbf{H}_j^\mathcal{H} \oplus \texttt{sort}( \{ \mathbf{H}_k^\mathcal{H} \mid (k, j) \in \mathcal{E}^\top \})   \right)  \ \ \forall j \in \texttt{TopologicalOrder}(\mathcal{E}^\top).
\label{eq:H_j_update2}
\end{align}
%For nodes with types not contained in $\mathcal{H}$,
The array $\mathbf{H}^\mathcal{H}$ provides two benefits. First, it uniquely identifies
the (sub)query pattern when including \underline{only} the features in $\mathcal{H}$, used below to define graph-level string $h^{\mathcal{H}} \in \{0, 1\}^{256}$. Second, it establishes a canonical ordering $\pi^\mathcal{H}$ on $\mathcal{V}$. The hash of a (sub)query pattern (given $\mathcal{H}$) is defined as:
\begin{equation}
    h^\mathcal{H} = \textrm{\Large \$}\left( \bigoplus_{j \in \pi^{\mathcal{H}}} \mathbf{H}_j^\mathcal{H} \right),  \ \ \ \  \ \ \ \textrm{with} \ \ \ \ \ \pi^{\mathcal{H}} = \arg\textrm{sort}(\{ \mathbf{H}_j^\mathcal{H} \}_{j \in \mathcal{V}}).
\label{eq:h_and_pi}
\end{equation}

\textbf{Feature Extraction.}
Our framework allows configuring feature extractors,
each extractor function $f : \{0, 1\}^* \rightarrow \mathbb{R}^{d_f}$ converts string features for one node, into a numerical vector of $d_f$ dimensions.
We program simple feature extractors that we list in Appendix \ref{sec:appendix_feature_extractors}.
We now introduce our most-important object.
Let feature vector $\mathbf{x}^{\mathcal{H}}_{\mathcal{F}}$ contain features of nodes extracted from graph using $\mathcal{F}$, while using the canonical node ordering induced by $\pi^\mathcal{H}$. Formally:
\begin{equation}
\mathbf{x}^{\mathcal{H}}_{\mathcal{F}} = \bigoplus_{j \in \pi^\mathcal{H}} \left\{ f_{(t, a)}(\mathcal{X}_j[(t, a)])   \ \ \mid \ \ t = \tau_j \right\}_{(t, a) \in \mathcal{F}}.
\label{eq:x}
\end{equation}
For completeness, the dimensionality of $\mathbf{x}^{\mathcal{H}}_{\mathcal{F}}$ is given by $\sum_{(t, a) \in \mathcal{F}} \sum_{j \in \mathcal{V}} \mathbf{1}_{[t = \tau_j]} d_{f_{(t,a)}}$.
%However, this information it optional for the remainder of the paper.
It is important to note that the dimensionality of $\mathbf{x}^{\mathcal{H}}_{\mathcal{F}}$'s from two different (sub)query graphs, will be equal if the two graphs have the same number of nodes for every node type $t \in \mathcal{T}$. Theorems \ref{thm:canonical_feature}\&\ref{thm:cross_learning} have details.

Objects $\mathcal{F}$ and $\mathcal{H}$ are configurations and not functions of any particular query graph $\mathcal{G}$. In contrast, the objects $\mathbf{x}^\mathcal{H}_\mathcal{F}$, $\pi^\mathcal{H}$, $\mathbf{H}^\mathcal{H}$, and $h^\mathcal{H}$ are functions of the input $\mathcal{G}$ and should've written as $\mathbf{x}^\mathcal{H}_\mathcal{F}(\mathcal{G})$, \textit{etc}.

\begin{figure}[!h] \centering
\begin{minipage}{\linewidth}
\includegraphics[width=\linewidth]{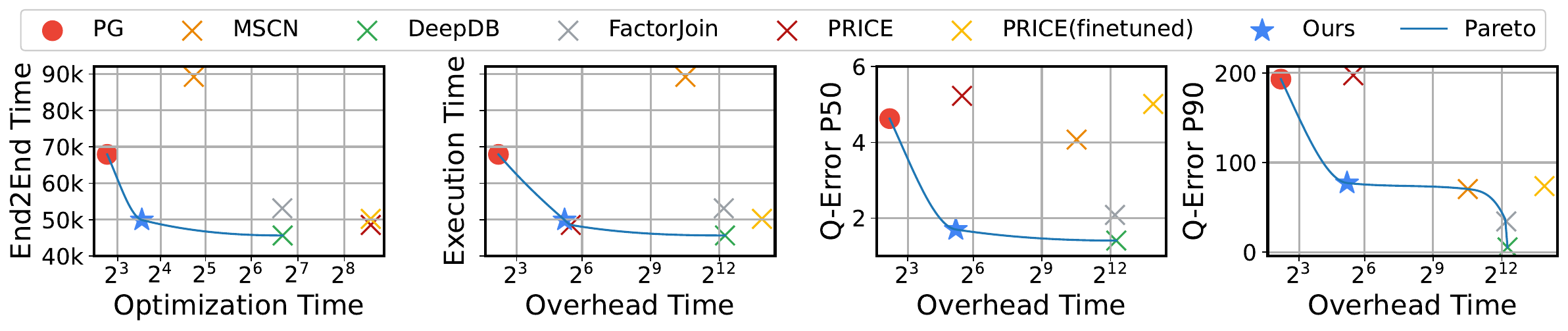}
\vspace{-0.7cm}
\captionof{figure}{Comparing different techniques on the IMDb database on multiple metrics.
  %including End-to-End time, total optimization time, total overhead time (e.g., time spent on training), and Q-Errors.
  Lower and to the left is better. Note the x-axis log scale.}
  \label{fig:scatter}
\end{minipage}

\vspace{0.1cm}

\begin{minipage}{.45\linewidth}
%\vspace{-0.1cm}
  \centering
\includegraphics[width=\linewidth]{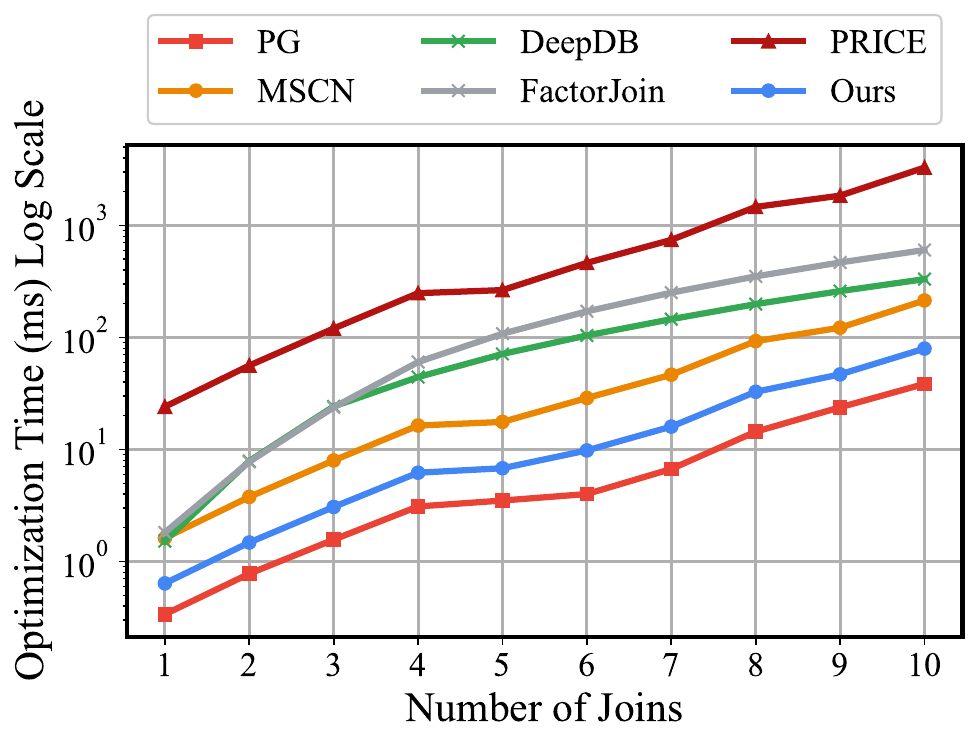}
    \vspace{-0.7cm}
    \captionof{figure}{Query Optimization Time Comparison per query on the IMDb dataset. Note the log scale on the Y-Axis.}
\label{fig:overhead}
\end{minipage}%
\begin{minipage}{.1\textwidth}
$\ $
\end{minipage}%
\begin{minipage}{.45\textwidth}
%\vspace{-0.7cm}
  \centering
    \includegraphics[width=\linewidth]{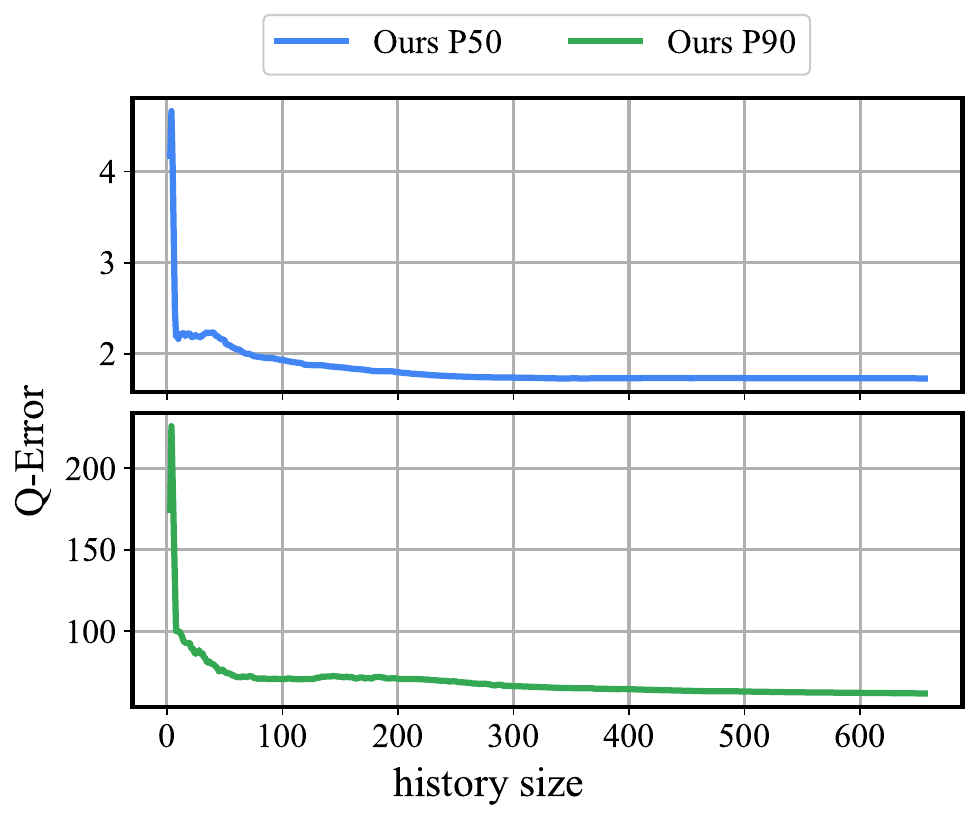}
    \vspace{-0.7cm}
    \captionof{figure}{Comulative Q-Error percentile on the IMDb workload VS size of set $\mathcal{D}^\mathcal{H}_\mathcal{G}$ (\S \ref{sec:online_algorithm}) }%Eq.\ref{eq:g_historysize}.}
    \label{fig:history_size}
\end{minipage}

\vspace{0.1cm}

% \end{figure}
% \begin{figure}[t]
% \centering
\begin{minipage}{.45\linewidth}
  \centering
   \includegraphics[width=\linewidth]{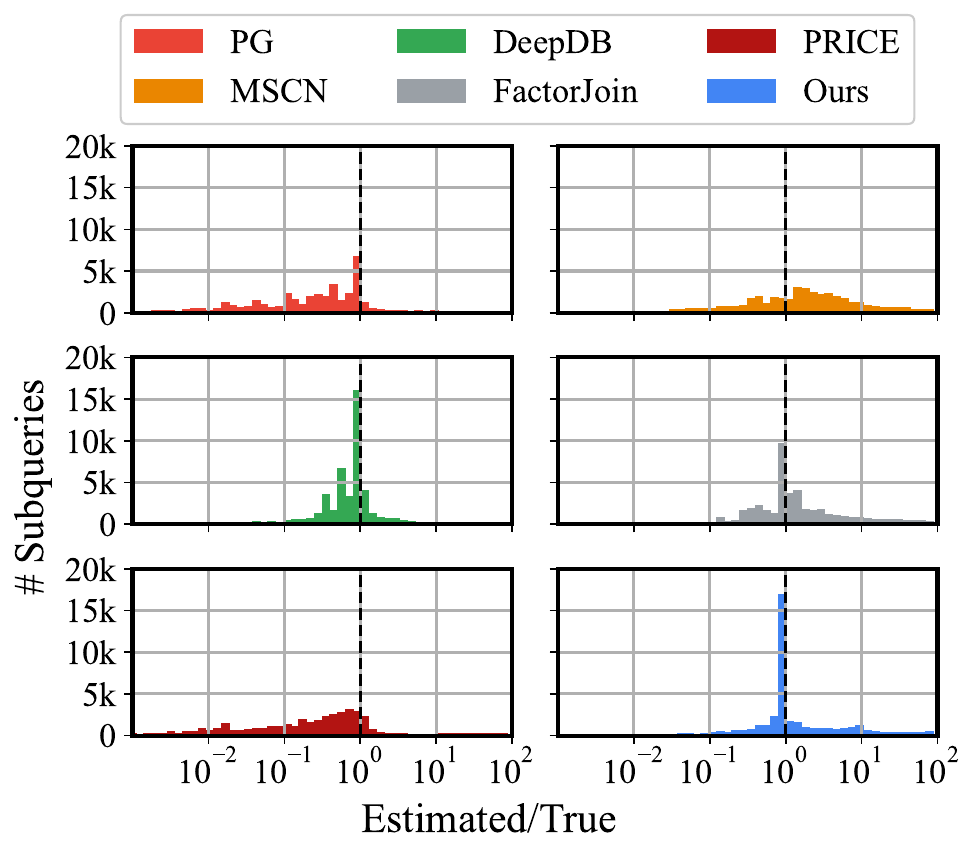}
   \vspace{-0.7cm}
    \captionof{figure}{Relative Estimation Errors Histogram on all 46,928 subqueries of IMDb. %workload.
    }
   \label{fig:q-error}
\end{minipage}%
\begin{minipage}{.1\textwidth}
$\ $
\end{minipage}%
\begin{minipage}{.45\textwidth}
  \centering
    \includegraphics[width=0.95\linewidth]{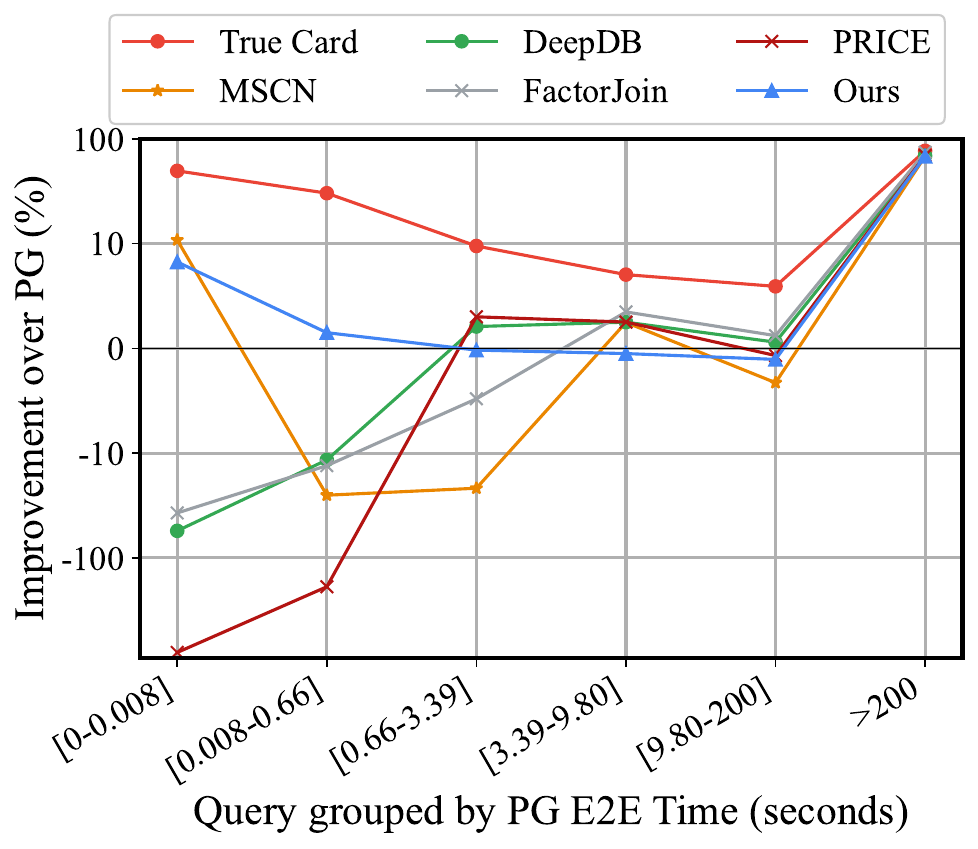}
    \vspace{-0.3cm}
    \captionof{figure}{Relative E2E time improvement over PostgreSQL by runtime group. $>$0 means improvements.
    %, while $<$0 means regression.
    %\ours performs robustly across runtimes, avoiding short query degradation while achieving gains on long ones.
    }
\label{fig:detailed_runtime}
\end{minipage}

\vspace{0.1cm}

% \end{figure}
% \begin{figure}[t]
% \centering
\begin{minipage}{.45\linewidth}
  \centering
  \includegraphics[width=0.85\linewidth]{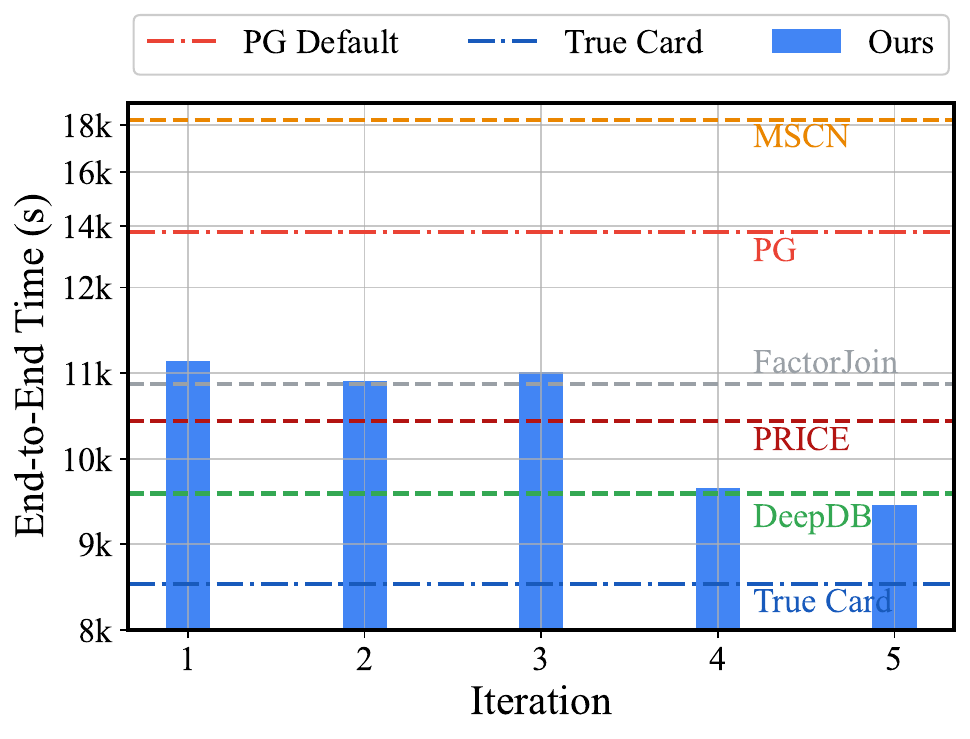}
  \vspace{-0.3cm}
  \captionof{figure}{
  E2E on IMDb. Runtime continuously improves  relative to static baselines.
  }
   \label{fig:multi_iteration}
\end{minipage}%
\begin{minipage}{.1\textwidth}
$\ $
\end{minipage}%
\begin{minipage}{.45\textwidth}
  \centering
  \includegraphics[width=0.85\linewidth]{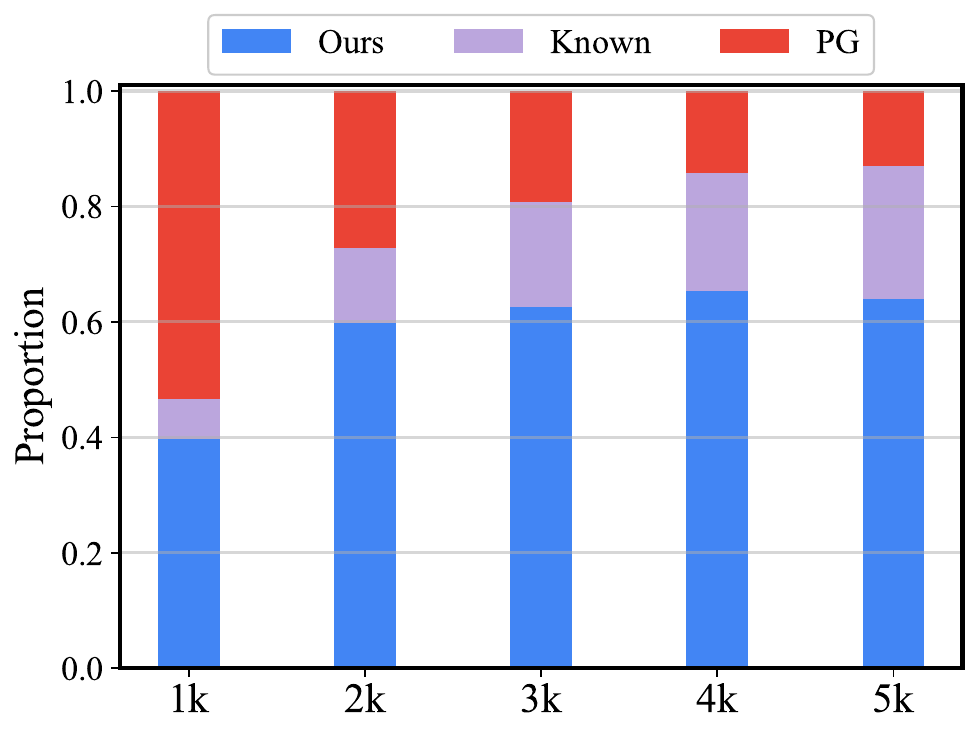}
  \vspace{-0.3cm}
  \captionof{figure}{
  Proportion of reliance on our models VS Postgres as history $\mathcal{D}$ accumulates while simulating the 5k IMDb workload.
  }
    \label{fig:proportion}
\end{minipage}
\end{figure}

\subsection{Correctness Analysis}
\vspace{-0.3cm}
We establish three theorems and present their ideas. The first two guarantee consistency within any graph, while the last enables learning across graphs. Formal theorems and proofs are in Appendix~\ref{appendix:correctness_proofs}. 

\textbf{Theorem Idea 1}\ \textit{
    Any feature set $\mathcal{H} \subseteq \mathcal{A}$ can induce a canonical node ordering.
}\newline    
\textbf{Theorem Idea 2}\ \textit{
    The sets $\mathcal{H} \subseteq \mathcal{A}$ and $\mathcal{H} \subseteq \mathcal{A}$ can extract a canonical feature vector.
}\newline
\textbf{Theorem Idea 3}\ \textit{
    Given an arbitrary anchor graph $\mathcal{G}$, then every $\mathbf{x} \in \{ \mathbf{x}^\mathcal{H}_\mathcal{F}(\mathcal{G}') \mid h(\mathcal{G}) = h(\mathcal{G}')  \}$ has the same dimensionality, with canonical node-to-feature positions.
}

%are inefficient, requiring accessing entire history for each inference, however, 

% We implement $K^\mathcal{H}(G, G')$ referenced in Eq.\ref{eq:localmodel}\&\ref{eq:oneshot} by plugging $h^\mathcal{H}$ of Eq.\ref{eq:x} into indicator function:
% \begin{equation}
% K^\mathcal{H}(G, G') = \mathbf{1}_{(h^\mathcal{H}(G) = h^\mathcal{H}(G'))}.
% \label{eq:kH}
% \end{equation}
% We define $g$ only in terms of features ignoring graph structure, i.e., $g(\mathcal{G}) \triangleq g(\mathbf{x}^\mathcal{H}_\mathcal{F}(\mathcal{G}))$
% and implement $K^\mathcal{H}_\mathcal{F}(G, G')$ as similarity among $\mathbf{x}^\mathcal{H}_\mathcal{F}(\mathcal{G})$ and $\mathbf{x}^\mathcal{H}_\mathcal{F}(\mathcal{G}')$.

% Our choice of Eq.~\ref{eq:kH} allows avoiding summing over $\mathcal{D}$ in Eq.\ref{eq:localmodel}\&\ref{eq:oneshot} and instead

% In addition, it provides

% In particular, let $\pi^{\mathcal{H}}$ be the topological ordering of $\mathcal{V}$ induced by $\mathcal{E}$. Since $\mathcal{E}$ represents a DAG, then \textbf{at least one} topological ordering must exist. We use $\mathbf{X}^\mathcal{H}$ to break ties

% have the same \textit{structure} and have identical values for every attribute in $\mathcal{H}$, then they will \textcolor{red}{interact}

% \subsection{Binary Kernel via Hashing}

% Let $\mathcal{\pi}_{\mathcal{H}}(G)$ define an ordering on $\mathcal{V}$, as defined by pattern features $\mathcal{H}$. Topological ordering breaking ties according to $\mathcal{H}$ The ordering is useful 

\subsection{Efficient Online Algorithm}
\label{sec:online_algorithm}
\vspace{-0.3cm}
Inference on test $\mathcal{G}$
\textit{seems} inefficient due to summation over history $\mathcal{D}$ (Eq.~\ref{eq:localmodel} \& ~\ref{eq:oneshot}), however,
our choice of $K^{\mathcal{H}}_{\mathcal{F}}$ (Eq.~\ref{eq:kernel}) allows random-access lookup
of $\{(\mathcal{G}', y) \mid h^{\mathcal{H}}(G) = h^{\mathcal{H}}(G')\}_{(\mathcal{G}', y) \in \mathcal{D}} \triangleq \mathcal{D}^\mathcal{H}_\mathcal{G}$.
In particular, we store in-memory $\texttt{HashTable } : \ \  h^{\mathcal{H}}(G) \ \ \ \mapsto \ \ \    \{ (\mathbf{x}^\mathcal{H}_\mathcal{F}(\mathcal{G}'), y') \}_{(\mathcal{G}', y') \in \mathcal{D} }$.
%a hashtable where \textit{key} $\mapsto$ \textit{value} are as:
%\begin{equation}
%
%\end{equation}
In fact, we never keep $\mathcal{D}$ in memory. After subquery $\mathcal{G}$ is executed, we append its feature vector $\mathbf{x}^\mathcal{H}_\mathcal{F}(\mathcal{G})$ and its cardinality onto $\texttt{HashTable}[h^{\mathcal{H}}(G)]$ then discard $\mathcal{G}$ to reduce memory footprint.
It is possible to further improve the efficiency in multiple ways. For instance, avoid frequent model fitting for $g^\textrm{DF}$ and $g^\textrm{LR}$ (Eq.\ref{eq:localmodel}), \textit{e.g.}, by storing model parameters, or use approximate nearest neighbors for $g^\textrm{RBF}$ (Eq.\ref{eq:oneshot}). However, further optimizations are outside the context of this paper, as our setup suffices for our experiments,
already speeding IMDb 5k workload by $>$7 minutes faster
with negligible total overhead time of $<$40 seconds. 

\subsection{Hierarchical Data Structure}
\label{sec:hierarchy}
\vspace{-0.3cm}
\begin{table}[t]
\caption{Features used for hashing and model invocation. The choices $\mathcal{H}_1 \subset \mathcal{H}_2 \subset \mathcal{H}_3$
to divisively partition subqueries, forming a hierarchy, as depicted in Fig.~\ref{fig:hierarchy}.}
    \centering
    \vspace{-0.3cm}
    {
    \footnotesize
    %\begin{tabular}{c p{3.6cm} p{3.6cm}}
    \begin{tabular}{c l l}
    \toprule
    % \rowcolors{2}{gray}{white}
$k$ & $\mathcal{H}_k$  & $\mathcal{F}_k$ \\
\cmidrule(lr){1-1}\cmidrule(lr){2-2}\cmidrule(lr){3-3}
1 & $\mathcal{H}_1 = \{(table, name), (column, type)\}$ &  $ \mathcal{F}_1 = \mathcal{F}_2 \cup  \{ (column, numUniques) \} $  \\
2 & $\mathcal{H}_2 = \mathcal{H}_1 \cup \{ (column, name) \}$  & $ \mathcal{F}_2 = \mathcal{F}_3 \cup  \{ (op, code) \} $ \\
3 & $\mathcal{H}_3 = \mathcal{H}_2 \cup \{ (op, code) \}$  & $\mathcal{F}_3=\{(literal, value)\}$  \\
\bottomrule
\end{tabular}
}
        \label{tab:template}
\end{table}
\begin{figure}[t]
    \centering    \includegraphics[width=0.8\linewidth]{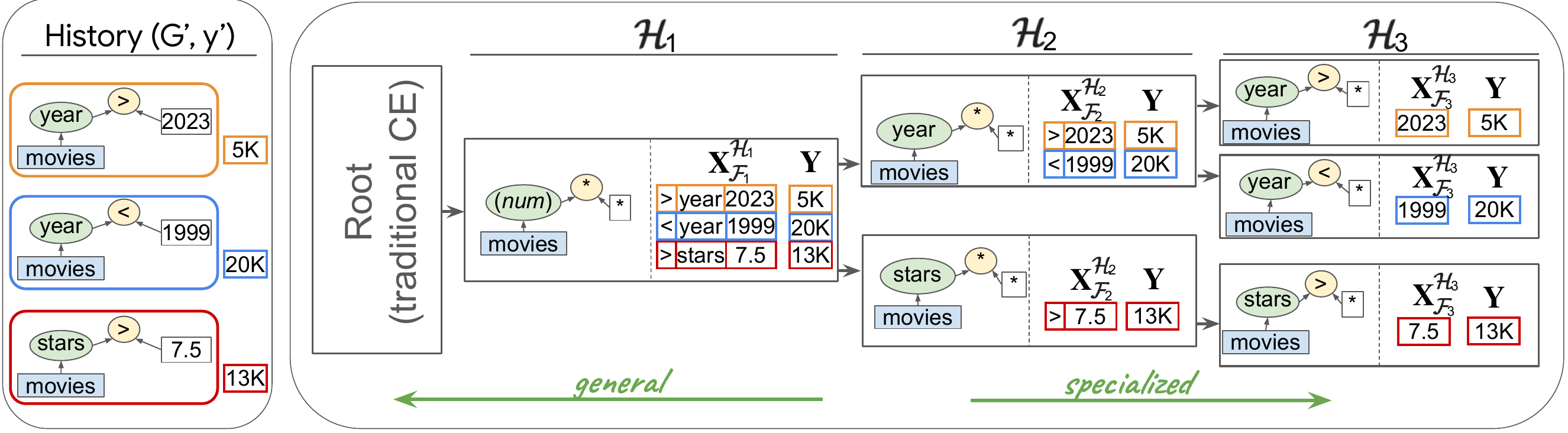}
    \caption{\textbf{(left)} Subqueries and their cardinalities arrive online, and get stored onto a \textbf{(right)} \texttt{HashTable} whose entries are keyed by (hash of) graph pattern, and the values are features extracted from graphs matching the pattern.
    The entries can be arranged as a hierarchy. Inference on test graph $\mathcal{G}$ walks the hierarchy from-right-to-left.
    If HashTable stores many observations under key $h^{\mathcal{H}_3}(\mathcal{G})$, then the entry's values will be used for inference. If there are only few observations, then the process is repeated with $\mathcal{H}_2$, \dots, falling-back onto heuristic %(``\textit{traditional}'')
    cost-estimator
    %(e.g., of PostgreSQL)
    for novel patterns.
    }
    \label{fig:hierarchy}
\end{figure}

Rather than one choice for each of $(\mathcal{H}, \mathcal{F})$, we include three $\{(\mathcal{H}_1, \mathcal{F}_1), (\mathcal{H}_2, \mathcal{F}_2), (\mathcal{H}_3, \mathcal{F}_3) \}$ and particularly choose
$\mathcal{H}_1 \subset \mathcal{H}_2   \subset  \mathcal{H}_3$, as listed in Table~\ref{tab:template}.
The choice of $\mathcal{H}'s$ recursively partitions subqueries into a hierarchy of three levels, yielding a data-structure depicted in ~\ref{fig:hierarchy}.
$\mathcal{H}_1$ is the most general.
As visualized in Fig.~\ref{fig:hierarchy}, $h^{\mathcal{H}_1}$ hashes subquery graphs to the same hash value, even though they differ on the op-code or the column name. Then, $h^{\mathcal{H}_2}$ partitions those by column. Finally, $h^{\mathcal{H}_3}$ partitions those by op-code.
For inference, we trust the most-specialized model with sufficient observations. Specifically, if $|\mathcal{D}^{\mathcal{H}_3}_\mathcal{G}| \ge \beta_3$, then inference is done using the model associated with $\texttt{HashTable}\left[h^{\mathcal{H}_3}(G)\right]$, else if $|\mathcal{D}^{\mathcal{H}_2}_\mathcal{G}| \ge \beta_2$, then using  $\texttt{HashTable}\left[h^{\mathcal{H}_2}(G)\right]$, else if $|\mathcal{D}^{\mathcal{H}_1}_\mathcal{G}| \ge \beta_1$, then using  $\texttt{HashTable}\left[h^{\mathcal{H}_1}(G)\right]$, else, then using the traditional cost estimator.

% We now describe our online learning system. Given a query ``\texttt{SELECT ... FROM ... WHERE ...}'',
% the query planner will seek cardinality estimates for the query itself, as will as its (possibly many) subqueries.
% Let $\mathcal{G}$ correspond to the graph of the query (or a subquery), depicted in Fig.~\ref{fig:annotated_sql}. Also, let
% $\mathcal{D} = \{ (\mathcal{G}_1, y_1), (\mathcal{G}_2, y_2), \dots \}$ denote historical (sub)query graphs and their corresponding ground-truth cardinalities $y$'s. At initialization, $\mathcal{D}$ starts empty. \textcolor{red}{Use superscripts for G and y?}

% When estimating the cardinality of

\section{Experimental Evaluation}
\label{sec:expr}
This section presents the main results. Appendix \ref{appendix:experiments} contains more experiments and discussions.
We evaluate cardinality estimation errors and impact on query execution time by investigating:
\begin{enumerate}[topsep=0pt, itemsep=0pt, leftmargin=15pt]
    \item How does \ours's performance (End-to-End time, accuracy) balance with its practical costs (optimization, training overhead), positioning it on the practical Pareto frontier? %(\S\ref{sec:expr_perf_overhead})
    \item A detailed analysis of \ours's performance, including runtime improvement for different groups, estimation error distribution, and gains from online learning. %(\S\ref{sec:expr_detail}).
    \item How do core design choices impact \ours's effectiveness? %(\S\ref{sec:expr_ablation})
\end{enumerate}

\textbf{Datasets and Workloads.} 
\begin{table*}[t]
\centering
\caption{Workload Stats. IMDb is from \citet{leis2015good} and others are from \citet{cardbench}}
\label{tab:workload}
\footnotesize
\vspace{-0.3cm}
\begin{tabular}{@{}r|cccc|ccc@{}}
\toprule
Workload                                   & Tables & Columns & Rows & Join Paths & Queries & Joins & Templates \\ \midrule
IMDb                                       & 6         & 37         & 62M     & 15            & 4972       & 1-4      & 40           \\ \midrule
stackoverflow & 14        & 187        & 3.0B    & 13            & 16,000     & 1-5      & 1440         \\
airline       & 19        & 119        & 944.2M  & 27            & 20,000     & 1-5      & 1400         \\
accidents     & 3         & 43         & 27.4M   & 3             & 29,000     & 1-2      & 1450         \\
cms           & 24        & 251        & 32.6B   & 22            & 14,000     & 1-5      & 2380         \\
 geo           & 16        & 81         & 8.3B    & 15            & 13,000     & 1-5      & 780          \\
 employee      & 6         & 24         & 28.8M   & 5             & 62,000     & 1-5      & 2480         \\ \bottomrule
\end{tabular}
\vspace{-0.1cm}
\end{table*}
We evaluate \ours using the IMDb dataset~\citep{leis2015good} and various workloads from CardBench~\citep{cardbench}. 
Table~\ref{tab:workload} summarizes dataset statistics. The IMDb dataset comprises $\approx$ 5000 queries derived from 40 JOB-Light\footnote{https://github.com/andreaskipf/learnedcardinalities/blob/master/workloads/job-light.sql} templates, used for overall performance and overhead evaluation. CardBench datasets, featuring queries with up to 5 joins, conjunctions, disjunctions, and string predicates, are primarily used for ablation studies and demonstrating generality, as many baselines lack support for these complexities, \textit{e.g.}, DeepDB, MSCN, PRICE lack string predicates and disjunction support.  % ; FactorJoin is limited to IMDb. -- limited in what way?

%\sparagraph{Datasets and Workloads.} We use two sets of benchmarks to evaluate our model: 1) the IMDb dataset ~\cite{leis2015good} and further extend the orignial 70 JOB-Light\footnote{https://github.com/andreaskipf/learnedcardinalities/blob/master/workloads/job-light.csv} queries to 5000 queries. These 5000 queries retain the same 40 query templates as the original JOB-Light queries, with only the constant values being modified; 2) the CardBench datasets and workloads ~\cite{cardbench}. Notably, the CardBench workload introduces increased complexity, featuring queries with up to 5 joins, conjunctions, disjunctions, and string predicates. However, since most techniques do not support these more complex predicates and datasets, we only report our Q-Error improvements from \postgres in the ablation studies section.

\begin{table*}[t]
\centering
\caption{Total End-to-End (E2E) Time, Total Overhead Time and Q-Error Performance Comparison for the 5k JOB-Light queries on the IMDb Database. E2E = Execution + Optimization.}
\label{tab:performance}
\footnotesize
\vspace{-0.3cm}
\begin{tabular}{@{}cccccccc@{}}
\toprule
\multirow{2}{*}{} &  \multicolumn{4}{c}{Runtime (in seconds)} & \multicolumn{3}{c}{Q-Error} \\
\cmidrule(l){2-5}  \cmidrule(l){6-8} 
                  &   E2E              &       Execution                &    Optimization                  &        Overhead           & P50    & P90      & P95     \\ \midrule
\postgres        & 67902                             & 67895                            & 6.72                                   &4.20                                  & 4.63   & 193.00   & 948.15  \\
\true         & 40476                             & 40476                            & /                                      & /                                  & 1.00   & 1.00     & 1.00    \\ \midrule
\mscn              & 89194                             & 89167                            & 26.77                                  & 1466.28                            & 4.07   & 70.39    & 219.31  \\
\deepdb            & 45635                             & 45532                            & 102.27                                  & 4860                               & 1.41   & 5.31     & 11.98   \\
\factorjoin        & 53095                             & 52994                            & 101.69                                  & 4680                               & 2.08   & 34.26    & 92.99   \\
\price             & 48520                             & 48142                            & 378.54                                 & 45.20                                   & 5.23   & 197.27   & 517.31  \\
\price(FT)  & 50190                             & 49812                            & 378.54                                 & 14828                              & 5.02   & 73.69    & 117.41  \\ \midrule
Ours              & 49895                             & 49883                            & 11.88                                   & 37.29                                 & 1.70   & 77.12    & 350.19  \\ \bottomrule
\end{tabular}
\vspace{-0.3cm}
\end{table*}

\textbf{System Setup.}
All experiments were conducted on a 64-Core AMD EPYC 7B13 CPU and 120GB RAM.
%  Linux machine equipped with a 
Like ~\citet{cardest}, we ran \postgres on a single CPU and disabled GEQO\footnote{https://github.com/Nathaniel-Han/End-to-End-CardEst-Benchmark}. %  (Genetic Query Optimizer)
%to assess the impact of cardinality estimation on query latency.

\textbf{Techniques.}
We compare \ours against default \postgres and representative state-of-the-art learned estimators across different paradigms: workload-driven (\mscn), data-driven (\deepdb, \factorjoin), and zero-shot (\price).
\begin{itemize}[topsep=0pt,itemsep=0pt,leftmargin = 15pt]
\item{\postgres ~\citep{postgres_estimator}. Denotes \postgres's  cardinality estimator.}
\item{\textsc{Oracle}. Emits the correct cardinality, establishing lower-bounds on errors and runtimes.}
\item \mscn ~\citep{mscn}: Multiset neural network that learns: query $\rightarrow$ cardinality. The model was trained using author-provided code for 200 epochs.
%  "the 100k queries provided in their code" ?

\item\deepdb ~\citep{deepdb}: data-driven approach that learns 
a sum-product network for each selected subset of tables in the database.

\item \factorjoin ~\citep{factorjoin}: a data-driven approach that applies factor graph on single tables and aggregates histograms for multiple tables.
\item \price ~\citep{price}: zero-shot approach, with parameters pre-trained on 30 datasets.
The overhead time for the base zero-shot model (45s in Table~\ref{tab:performance}) is incurred for computing necessary statistics such as histograms, fanout, common value counts, and table sizes.
%; we optimized the fanout computation from the authors' original Python implementation.

\item \price(FT) We fine-tuned the above, using their code-base, on 50k queries for 100 epochs.
\item{\ours}: Ours, following \S\ref{sec:online_algorithm} \& \S\ref{sec:hierarchy}, performs online learning, starting from scratch and incrementally refining models as new queries arrive. We set $\beta_3 = 10$, $\beta_2 = 50$, $\beta_1 = 100$.

% C++
%Training utilizes true cardinalities of the subqueries (subquery, cardinality) in the executed plans selected by \postgres. We implemented the templatization and hierarchical models in C++ code. More specifically, we use the YDF library ~\cite{ydf} for Gradient Boosted Decision Tree. We set $\tau_1 = 10, \tau_2 = 50, \tau_3 = 100$ in Section ~\ref{sec:online_learning}.
\end{itemize}

\textbf{Evalutaion Metrics.} We evaluate our proposed method against alternatives using error metrics and run-times. \textbf{Q-Error} metric \citep{moerkotte2010preventing}  quantifies the relative deviation of the predicted ($\hat{y}$) from the true cardinality ($y$). Lower is better, with 1 implying perfect estimation, defined as:
   \begin{equation}
       %Q_\textrm{err} =  \max\left(\frac{y}{\hat{y}}, \frac{\hat{y}}{y}\right)
       Q_\textrm{err} =  \max\left({y} /{\hat{y}}, \ {\hat{y}}/{y}\right)
   \end{equation}
    To understand both typical and tail estimation errors, we report Q-errors percentiles \{50, 90, 90\}. Further, and more importantly for the user, we report the following run times:
%\begin{itemize} [topsep=0pt, itemsep=0pt, leftmargin = 15pt]
    %\item
    \textbf{End-to-End (E2E)} query-to-response latency, measured by replacing cardinality estimation of PostgreSQL (v 13.1) with (aforementioned) alternative techniques, per work of
    \citet{cardest};
    %, as released in CardEst ~,
    %to measure the runtime when replacing 
    %\item
    \textbf{Optimization time} spent by the query optimizer to generate a plan, including the time to obtain cardinality estimates for all subqueries considered by the optimizer;
    %\item
    \textbf{Overhead time} required for training or updating the cardinality estimation model.
    For offline, data-driven or query-driven approaches, this is bulk training time. 
    For our online approach, this is the time for incremental updates.
    Note: we \textbf{do not} include the significant overhead of training data collecting  for query-driven methods, \textit{e.g.}, $\approx$ 34 hours for \mscn.
    %We report median (P50), 90th percentile (P90), and 95th percentile (P95) Q-errors 
%\end{itemize}

\subsection{Accuracy-Overhead Tradeoff: The Practical Pareto Frontier}
\label{sec:expr_perf_overhead}
\vspace{-0.3cm}
%\textcolor{red}{Our model can improve from PG default, both in end-2-end latency time and in q-errors. Show end2end time in IMDB dataset and q-errors in Imdb and other 6 more baselines.}

Achieving high estimation accuracy often comes at the cost of increased computation, creating a trade-off between accuracy (estimation and lower E2E time) 
and overheads (model updates and inference). Practical estimator should reside on the Pareto frontier in this multi-dimensional space.
% This sentence is repeated at end of section:
% Our experiments demonstrate that \ours occupies a unique and advantageous position on this frontier among the evaluated techniques, offering significant performance benefits with minimal operational costs.

%Table~\ref{tab:performance} and Figure~\ref{fig:scatter} summarize the key performance and cost metrics for all techniques on the 5000-query IMDb workload. Figure~\ref{fig:scatter} visually represents the trade-offs: ideally, a technique would be towards the bottom-left (low E2E Time, low Optimization Time, low Overhead with low Q-Errors).

% \begin{table}[t]
% \centering
% \caption{E2E Time and Q-Error Comparison on the IMDb Dataset.}
% \label{tab:job_light_performance}
% \begin{tabular}{l S[table-format=6.2] S[table-format=1.2] S[table-format=3.2] S[table-format=4.2]}
% \toprule
% Model & {Runtime (s)} & {$Q_\textrm{err}^{50}$} & {$Q_\textrm{err}^{90}$} & {$Q_\textrm{err}^{95}$} \\
% \midrule
% \postgres & 13118.04 & 4.59 & 209.09 & 1014.44 \\
% True Card & 8537.60 & 1.00 & 1.00 & 1.00 \\
% MSCN & 18236.57 & 4.46 & 73.33 & 233.78 \\
% DeepDB & 9594.39 & 1.44 & 5.03 & 10.75 \\
% FactorJoin & 10871.45 & 2.32 & 37.58 & 98.68 \\
% PRICE & 10441.94 & 4.83 & 193.14 & 516.64 \\
% PRICE (finetuned) & 10685.45 & 4.81 & 75.12 & 119.82 \\
% Ours, PG adjusted (Iter 0) & 11063.61 & 3.15 & 94.58 & 412.17 \\
% Ours, PG adjusted (Iter 1) & 9900.45 & 1.09 & 30.44 & 144.92 \\
% \bottomrule
% \end{tabular}
% \end{table}

\textbf{Overall Performance and Efficiency Comparison.}
Table~\ref{tab:performance} and Figure~\ref{fig:scatter} compares performance (End-to-End Time, Q-Error) and cost (Optimization Time, Training Time) across  all techniques on the 5k IMDb workload.
%We use all three $H_1, H_2, H_3$ and GBDT for the experiments.
%Table~\ref{tab:performance} provides numerical results, while Figure~\ref{fig:scatter} visualizes the multi-dimensional trade-offs.
%
%Analyzing the results, we see distinct clusters of techniques in the performance-efficiency space. 
We make the following obervations.
\begin{itemize}[topsep=0pt,itemsep=0pt,leftmargin = 9pt]
\item Default \postgres offers minimal optimization time (6.72s) and overhead time (4.20s) where the overhead time is the time running ANALYZE on the database.

\item Data-driven methods (\deepdb and \factorjoin) achieve significantly better Q-Errors (P90 5.31, 34.26) and improved End-to-End times (45635s, 53095s). However, this performance comes at the expense of substantially higher optimization times (102.27s, 101.69s) and massive training overheads (4860s, 4680s), representing a significant practical barrier.

\item Query-driven method \mscn achieves better Q-Error than \postgres (P50 4.07 vs 4.63, P90 70.39 vs 193), but paradoxically results in a worse End-to-End time - increased by from 67902s to 89194s (31\% degrade in performance).

\item Zero-shot approach \price achieves an End-to-End time of 48520s, an improvement over \postgres (67902s). However, it incurs a very high optimization time of 371.73s for the 5k query workload, significantly higher than both \postgres (6.72s) and \ours (11.88s). Base \price also exhibits higher Q-errors (P50 5.23, P90 197.27) compared to \ours (P50 1.70, P90 77.12) and the data-driven baselines. A fine-tuned version of \price, trained on a specific workload, improves Q-errors (P50 5.02, P90 73.69) but results in a slightly worse End-to-End time (50190s) and introduces a substantial training overhead of 14828s (over 4 hours) using CPU. This highlights that while fine-tuning can improve accuracy, it does not guarantee better End-to-End performance and introduces significant retraining costs.

\item \ours achieves a substantial 27\% reduction in End-to-End time (49895s vs 67902s) and significantly improves Q-errors (P50 1.70 vs 4.63, P90 77.12 vs 193.00). Crucially, it does this while maintaining an optimization time (11.9s) comparable to \postgres and incurring a negligible training overhead (37.3s total for the 5k query workload) than any other learned method. 
\end{itemize}

% \textcolor{red}{Show the overhead time comparison with other baselines. Show the benefits of our templatized model - easily updatable and plugged-in. }

\textbf{Optimization Time Scalability.}
Figure~\ref{fig:overhead} shows that cardinality estimation time \textbf{scales exponentially} with query complexity (number of joins).
Therefore, practical cardinality estimators must exhibit minimal latency.
The figure shows that default \postgres starts with low optimization time ($\approx 0.3$ ms for 1 join) and increases gradually. \ours mirrors this behavior, remaining comparable to \postgres across all join counts (\textit{e.g.}, $\approx 60-80$ ms at 10 joins), which is feasible because our lightweight models enable per-subquery estimates in $\approx 0.1$ ms.
In contrast, other baslines slow optimization by 10X-100X, posing a major practical barrier.

\section{Related Work}
\begin{table*}[t]
\caption{Summary of existing cardinality estimation approaches. Overhead is the initial setup cost for a new database. Optimization time is per-query cost. Updatability reflects responsiveness to workload/data shift.
Performance indicates end-to-end query latency.
%See Section \ref{sec:expr} for detailed measurements.
}
\label{tab:background}
\centering
\vspace{-0.3cm}
{\footnotesize
\begin{tabular}{@{}rlp{1.4cm}ll@{}}
\toprule
                     & New DB Overhead    & Infer Time (per query) & Updatability                              & Performance \\ \midrule
Traditional          & None                     & $0.1ms$                   & Fast                        & Moderate \\
Query-driven         & High (Collect \& Train) & $1ms$              & Slow, Batch Retrain           & Variable ($-$) \\
Data-driven          & High (Train on Data) & $1$--$10ms$                  & Slow, Retrain on Data Update & Good (++)\\
Zero-shot            & Low (Pre-trained)& $1$--$20ms$                  & Slow, Batch Finetune           & Good (+) \\
\midrule
\ours & None (Learn from History)         & $0.2ms$                   & Fast, Incremental        & Good (+) \\
\bottomrule
\end{tabular}
}
\vspace{-0.3cm}
\end{table*}

% Our work is most similar to \citet{blackbox-cardinality} where queries are grouped by template and a simple model is learned per template. However, we differ in three main ways: (1) our method predicts cardinalities for all subqueries (not just the main query!). (2) We establish a family of hash functions \ref{sec:ourmethod_implementation} that allows subqueries accessing different tables (or columns) to learn from one another. (3) Lastly, our template-grouping is based on the graph structure, and therefore is invariant to many textual transformations including aliases and ordering of conjunctions and tables.

% \paragraph{Template-based.}
% \url{https://dl.acm.org/doi/pdf/10.1145/3588963} also groups query instance parameters by query template. However, they assume that templates are given through an outside process. On the other hand, we automatically extract templates and corresponding query parameters by high-level configuration of $\mathcal{H}$'s and $\mathcal{F}$'s.

%We review related work, using two prospectives:  \textbf{Cardinality Estimation} and \textbf{Graph Algorithms}.

%\textbf{Cardinality estimation.}
%Accurate cardinality estimation, the process of predicting the number of rows returned by a query without running the query, is a cornerstone of effective query optimization. Errors in cardinality estimates can lead to suboptimal query plans, resulting in significantly increased query execution time. Here 
Table~\ref{tab:background} compares categories of cardinality estimators, detailed as follows.
\textbf{Traditional} techniques~\citep{postgres_estimator,url-mysql,adaptive_sampling,index_sampling}, such as histogram-based methods and sampling-based approaches, rely on simplified assumptions about data distributions and attribute independence.
While efficient and easily updatable, they often struggle with complex query patterns involving multiple joins,
%intricate predicates (e.g., range queries, LIKE predicates),
and correlated data, leading to large estimation errors.
\textbf{Query-driven methods}
%Query-driven approaches
frame cardinality estimation as a supervised learning problem, training models to map featurized query to cardinality -- \textit{e.g.}, feed-forward networks ~\citep{mscn,jpgm}, gradient boosted trees~\citep{lightweight_model}, and tree-LSTM~\citep{treelstm}.
These methods require training data \textbf{upfront} (rather than online) \textit{i.e.}, simulating and executiing queries while recording their cardinalities.
Training may be repeated when database contents or workloads shift.
%These approaches have relatively low optimization time.
%
Further, they add an overhead during query planning (inference) (\S\ref{sec:expr_perf_overhead}).
Our method is also supervised, though learns many simple models, online, one model per subquery pattern.
Our style of pattern-based learning had appeared earlier, \textit{e.g.}, \citep{blackbox-cardinality}, %since they also group queries by pattern, and also perform learning-and-inference on dense-vectors within each pattern.
however, we differ in: (1)
our patterns are graph rather than SQL text, which are invariant to aliases and ordering (e.g., of junctions); and (2) learning hierarchy of models rather than a one-level partitioning.
\textbf{Data-driven Methods}
directly model the table data distributions ~\citep{deepdb,flat,factorjoin,lightweightgm,neurocard}. They generally produces effective estimates and results in good end-to-end time performance. However, they typically incur long training time, large model size and slow optimization time. Updating these models when the underlying data changes is also slow and often requires expensive re-training.
\textbf{Zero-shot Methods} aim to transfer knowledge learned from a diverse set of pre-trained databases to a new database without requiring database-specific training data ~\cite{price}.
 While promising for cold-start scenarios, these methods can still suffer from high optimization time. Furthermore, while they can be fine-tuned on database-specific queries, this process can still be slow.

% \textbf{Graph Algorithms.} Our we draw similarities against other graph algorithms.
% \textbf{Hashing of DAGs}
% \citet{directed-graph-hashing} also extend Merkle Trees~\citep{merkle} to DAGs.
% Other methods that can operate on directed and undirected graphs, including \citet{graph-id}
% and WL \citep{wl-kernels}.
% These methods iteratively update node's hash using itself and its neighbors. Each update-round incorporates information from further neighbors.
% \textit{sometimes} (\textit{e.g.}, \textit{or} junction), while being variant at \textit{other times} (\textit{e.g.}, $>$ operator). In addition, we only work with DAGs and therefore iterating in topological order terminates the algorithm.
% %
% \textbf{Decoupled Graph Neural Nets.} Our method is also linked to methods that ``\textit{decouple}'' the graph-processing step from the learning. Specifically, methods that extract features using the graph and no longer need the graph for learning. These methods include \citep{simplegcn, sign}.
% In that regard, our method also uses the graph for pre-processing.
% We differ than those methods as they use the structure to propagate information along edges whereas we hash the structure.

% \subsubsection*{Acknowledgments}
% Use unnumbered third level headings for the acknowledgments. All
% acknowledgments, including those to funding agencies, go at the end of the paper.

\section{Conclusion}
We are interested in learning a cardinality estimator for diverse workloads.
Instead of a monolithic model that can handle any arbitrary query, we learn many simple models, each model specialized to one subquery pattern.
In particular,
we define cardinality estimation models
using a kernel function across Graphs.
The kernel deems two subqueries as similar if they are structurally-equivalent and they have similar features. 
Similar subqueries influence one another either when learning a local model (Eq.~\ref{eq:localmodel}) or with one-shot inference (Eq.~\ref{eq:oneshot}).
We presented an efficient implementation using an online learning algorithm
that extracts (feature-vector, cardinality) pair for every subquery graph,
and groups them by graph hash values.
Finally, we configure multiple hash functions and their corresponding learning features, such that, the query history can be recursively partitioned into a hierarchy.
The leaves of the hierarchy contain subqueries that are highly-similar
(\textit{e.g.}, equivalent, up-to constants and literals),
whereas first and intermediate levels of the hierarchy aggregate more general queries,
where nodes contain structurally-equivalent subqueries that read different columns
or use different op-codes.
Our method provides a uniquely compelling balance, achieving significant performance benefits and accuracy improvements over traditional methods with operational costs orders of magnitude lower than other learned techniques, positioning itself on the practical Pareto frontier for learned cardinality estimation.
% Experimentally,
% our method competes with other learning-based approaches in terms of cardinality estimates,
% and can train online with negligible overheads. This allows us to integrate with open-source PostgreSQL and measure the runtime improvement to be $>$30\% for IMDb 5K workload.

{\footnotesize
\bibliography{iclr2026_conference}
\bibliographystyle{iclr2026_conference}
}

\newpage
\section*{Appendix}
\appendix

\section{Directed Acyclic Graphs of SQL Queries}
\label{appendix:sql_to_dag}

We convert an input SQL query (\footnote{See Appendix for PostgreSQL's \texttt{RelInfo} data structure}) into a directed acyclic graph (DAG) in the following steps:
\begin{enumerate}[topsep=0pt, itemsep=0pt]
    \item Parse input statement as a parse-tree. It is possible to use an open-source parser, like \url{https://github.com/tobymao/sqlglot}. \label{item:parse}
    \item Merge identical nodes (column names or table names).
    \item For every referenced \textit{column}, we add two edges: Table $\rightarrow$ Table Alias\footnote{The alias is important as certain queries access one table twice, joining it with itself. Nonetheless, the alias name is ignored by our method.} $\rightarrow$ \textit{column}.
\end{enumerate}
The parse-tree (Step \ref{item:parse} above) 
already contains the predicate expression tree appearing in the ``\texttt{WHERE}''-clause, \textit{e.g.}, with nodes representing column names; operators (\texttt{=}, \texttt{>}, \texttt{+}, \texttt{not}, \dots); conjuctions and disjunctions (\texttt{and}, \texttt{or}); literals;  function names (\texttt{SUBSTRING}, \texttt{ABS}, \texttt{NOW}, \dots); etc.

\section{Integration with PostgreSQL}
\label{appendix:postgres_to_dag}

To evaluate the efficacy of \ours, we integrated it into open-source PostgreSQL as an extension, as depicted in Figure~\ref{fig:placeholder}. This integration involved adding new hooks into the PostgreSQL engine, enabling the query planner to utilize \ours for cardinality estimation, thereby influencing plan decisions and allowing the collection of performance statistics to demonstrate the efficacy of \ours approach. While this work focuses on demonstrating the core algorithm's efficacy, production-level optimizations such as memory management, storage and asynchronous training mechanisms are are beyond the scope of this paper.

\begin{figure}[h!]
    \centering
    \includegraphics[width=0.9\linewidth]{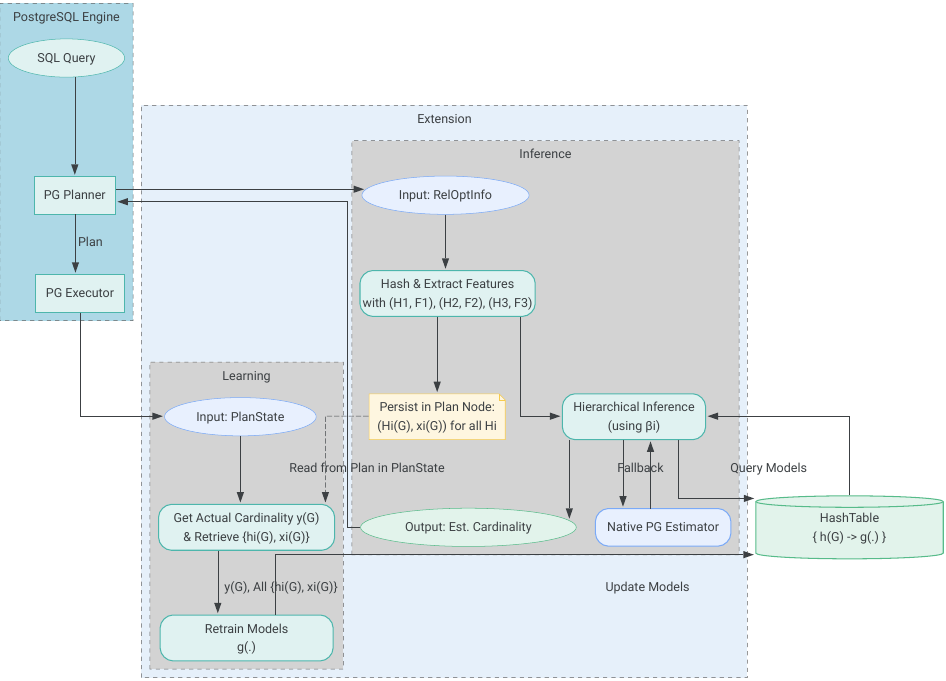}
    \caption{Integrating LITECARD  with PostgresSQL}
    \label{fig:placeholder}
\end{figure}

\subsection{Inference}
\ours interacts with the cost estimator at various points within the PostgreSQL planner to provide learned estimates. This is achieved using PostgreSQL's hook mechanism, specifically by setting hooks within functions such as \verb|set_baserel_size_estimates|
(PG cardinality estimation function for base relations) and \verb|get_parameterized_joinrel_size| (PG cardinality estimation function for join relations) and more. These hooks allow us to override the default cardinality estimates.
When the planner requires a cardinality for a relation (represented by \verb|RelOptInfo|), our hooks are invoked. We process the \verb|RelOptInfo| struct, analyzing filters (\verb|baserestrictinfo|), join information, and other plan attributes to generate hashes and corresponding features according to the strategies defined in \S\ref{sec:hash_and_feat}. 
The system attempts to predict cardinality using the model corresponding to $\mathcal{H}_3$. Following the hierarchical approach outlined in \S\ref{sec:hierarchy}, if the model for $\mathcal{H}_3$ does not meet the activation threshold $\beta_1$ (e.g., insufficient training samples), we fallback to the previous level in the hierarchy, $\mathcal{H}_2$, generating $h^{\mathcal{H}_2}(G)$ and $\mathbf{x}^{\mathcal{H}_2}(G)$ to invoke the corresponding $g(.)$. This process continues to $\mathcal{H}_1$ if necessary. If no model in the hierarchy is sufficiently confident, we fallback to the native PostgreSQL estimator, ensuring robustness.
The metadata generated during this process, 
including the hashes
$\left(h^{\mathcal{H}_1}(G), h^{\mathcal{H}_2}(G), h^{\mathcal{H}_3}(G)\right)$
and the extracted features
$\left(\mathbf{x}^{\mathcal{H}_1}(G), \mathbf{x}^{\mathcal{H}_2}(G), \mathbf{x}^{\mathcal{H}_3}(G) \right)$, and which hierarchical level provided the estimate, are persisted within the plan node structures (specifically within the Plan nodes). This information is crucial for online learning and observability.

\subsection{Learning}
The online learning mechanism (\S\ref{sec:ourmethod_graphlocal}) is realized through executor hooks. We use the \verb|ExecutorStart_hook| to ensure row count instrumentation is enabled for each node in the plan.
The \verb|ExecutorEnd_hook| is pivotal for capturing the ground truth after query execution.
Once execution is complete, for each node in the plan tree,
we retrieve the persisted hash value $h^{\mathcal{H}_i}(\mathcal{G})$ and features $\mathbf{x}^{\mathcal{H}_i}(\mathcal{G})$, along with the actual cardinality $y$ from the execution statistics. This triplet $\left(h^{\mathcal{H}_i}(\mathcal{G}), \mathbf{x}^{\mathcal{H}_i}(\mathcal{G}), y\right)$ constitutes a new training example. This example is used to update or retrain the parameters of the corresponding model $g(.)$, thus allowing the models to continuously adapt to the observed query workload. 

\subsection{Observability}
To facilitate understanding of \ours's behavior, we have enhanced the EXPLAIN ANALYZE command of PostgreSQL. The output for each plan node now includes the cardinality predicted by \ours, the inference latency for the \ours model, the hash $h^{\mathcal{H}_i}(\mathcal{G})$ used for the prediction, the features $\mathbf{x}^{\mathcal{H}_i}(\mathcal{G})$ extracted and the hierarchical level $i$ from which the prediction was made.

\subsection{Handling \postgres Bias}
\label{sec:bias}

Effectively integrating a learned estimator requires understanding and mitigating biases in the base optimizer. PostgreSQL's default estimator exhibits a significant underestimation bias, which can impede optimal plan selection.

\begin{table}[t]
\centering
\caption{PG (Biased) Cardinality Estimation Analysis on the IMDb database. Note that as the number of joins increases, the underestimate proportion and average Q-error increase drastically.}
\label{tab:pg_biased}
\begin{tabular}{ccc}
\toprule
{$n\_join$} & {Underestimate Proportion} & {Average Q-Error} \\
\midrule
1 & 0.57 & 1.57 \\
2 & 0.83 & 20.20 \\
3 & 0.93 & 1361.38 \\
4 & 0.98 & 68655.97 \\
\bottomrule
\end{tabular}
\end{table}

\sparagraph{\postgres 's Underestimate Bias.} 
Table~\ref{tab:pg_biased} quantifies the inherent underestimation bias in PostgreSQL's default cardinality estimates on the IMDb JOB-Light workload~\citep{leis2015good}. The table shows the proportion of subqueries underestimated by PostgreSQL and their average Q-error, grouped by join count. We observe the underestimation proportion sharply increases with joins (e.g., $>$80\% for 2-join, $>$98\% for 4-join queries). Correspondingly, average Q-error escalates dramatically, reaching over 68,000 for 4-join queries. This systematic underestimation is critical as optimizers rely on these estimates for plan choices; underestimates can lead PostgreSQL to select seemingly cheaper but suboptimal plans (e.g., favoring nested loops for intermediate results that are much larger than estimated).
Table~\ref{tab:pg_biased} demonstrates PostgreSQL's severe, join-dependent underestimation bias, a key factor leading to poor plan quality.

\sparagraph{Impact of Bias and Our Solution.}
\begin{figure}[t]
  \centering
    \centering
    \includegraphics[width=0.5\linewidth]{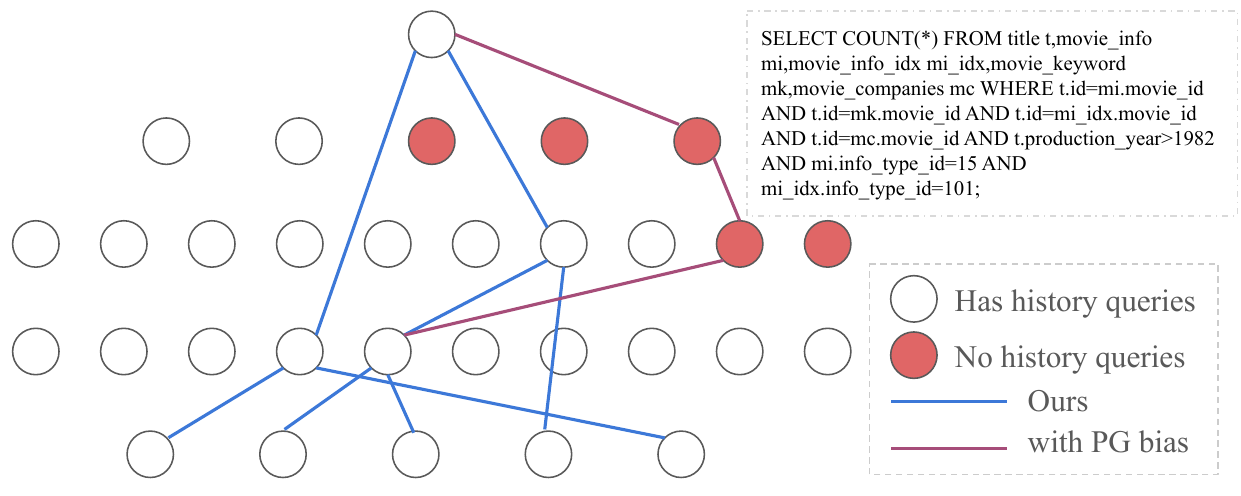}
  \caption{Query planning example illustrating the impact of PostgreSQL bias. Each node represents a subquery where the bottom level are the single table queries and the top node is the whole query.  Shows how an underestimate can lead to a disastrous plan path (3400s execution) and how adjusting the bias allows \ours to select a better plan (141s execution).}
\label{fig:bias_example}
\end{figure}
Figure~\ref{fig:bias_example} illustrates the impact of \postgres's bias using an example query from the 5000-query IMDb workload. If we naively combine estimates, \postgres's underestimate for subqueries lacking historical data (represented by the red nodes) leads to a disastrous plan executing in 3400 seconds. This occurs because \postgres's underestimate makes these subqueries appear smallest at their level, causing the optimizer to select them. 
To address this severe underestimate bias problem, we sample a probability number and then multiply their \postgres estimates by the average Q-errors documented in Table~\ref{tab:pg_biased}. For example, for a subquery at the third level involving 2 joins, we uniform sample a probability from 0 to 1, if it is smaller than 0.83 , we multiply the estimate by 20.2; for a fourth-level subquery involving 3 joins, if the sampled number is smaller than 0.98, we multiply by 1361.38. 
This bias information (\textit{e.g.} Table ~\ref{tab:pg_biased}) can be practically collected from executed queries for any database with minimal overhead. Figure~\ref{fig:bias_example} shows that applying this adjustment allows \ours to avoid the disastrous plan, resulting in a near-optimal execution time of 141 seconds, compared to PostgreSQL's default plan at 171 seconds and injecting true cardinality oracle at 133 seconds.

\section{Correctness Proofs}
\label{appendix:correctness_proofs}
\begin{definition} (Graph Isomorphism under feature set)
Let graphs $\mathcal{G}$ and $\mathcal{G}'$ be isomorphic under feature-set $\mathcal{H}$, denoted as \fbox{$\mathcal{G} \underset{\mathcal{H}}{\cong}\mathcal{G}'$} if-and-only-if there exists a bijection $\pi_{(.)} : \mathcal{V} \rightarrow \mathcal{V}'$
such that
\begin{equation}
 \mathcal{E}' = \{(\pi_{u}, \pi_v) \}_{(u, v) \in \mathcal{E}}
 \textrm{ \ \ \ \textbf{and} \ \ \  }
\mathcal{X}'_{\pi_j}[(t, a)] = \mathcal{X}_{j}[(t, a)]
\textrm { for all } (t, a) \in \mathcal{H} \textrm{ and }  j \in \mathcal{V} 
 \end{equation}
 \label{definition:isomorphism_feats}
\end{definition}

\begin{definition} (Predecessors)
    Let $\mathcal{P}_j \subset \mathcal{V}$ be the predecessors to node $j \in \mathcal{V}$ defined as follows. Given edge $(u, v) \in \mathcal{E}$, its starting-point $u$ will be included in $\mathcal{P}_j$ if either  $v =j$ or $v \in \mathcal{P}_j$.
\end{definition}
\begin{definition} (Successors)
    Let $\mathcal{S}$ the equals the $\mathcal{P}$ corresponding to the reverse graph $(\mathcal{V}, \mathcal{E}^\top, \mathcal{X})$.
\end{definition}

\begin{theorem}
    Any feature set $\mathcal{H} \subseteq \mathcal{A}$ can induce a canonical node ordering. Specifically,
    \begin{align}
    \mathcal{G} \underset{\mathcal{H}}{\cong} \mathcal{G}' \ \  &\Longrightarrow \ \
    \mathbf{A}^{\pi^\mathcal{H}(\mathcal{G})} \times  \mathbf{H}^\mathcal{H}(\mathcal{G})  = 
    \mathbf{A}^{\pi^\mathcal{H}(\mathcal{G}')} \times  \mathbf{H}^\mathcal{H}(\mathcal{G}')
    \label{eq:canonical_implication_1}   \\
    \mathcal{G} \underset{\mathcal{H}}{\cong} \mathcal{G}' \ \  &\underset{whp}{\Longleftarrow} \ \
    \mathbf{A}^{\pi^\mathcal{H}(\mathcal{G})} \times  \mathbf{H}^\mathcal{H}(\mathcal{G})  = 
    \mathbf{A}^{\pi^\mathcal{H}(\mathcal{G}')} \times  \mathbf{H}^\mathcal{H}(\mathcal{G}'), \label{eq:canonical_implication_2}
    \end{align}
such that $\pi^\mathcal{H}(\mathcal{G})$ and $\pi^\mathcal{H}(\mathcal{G}')$ can be used to align the featured DAGs,  and sparse re-ordering (adjacency) matrix $\mathbf{A}^{\pi^\mathcal{H}(\mathcal{G})} \in \{0, 1\}^{n \times n}$ shuffles rows of its multiplicand according to ordering defined by $\pi^\mathcal{H}(\mathcal{G})$, as:
\begin{equation}
    A_{i,j}^{\pi^\mathcal{H}(\mathcal{G})} = \mathbf{1}_{\left[j \  = \  \pi_i^{\mathcal{H}}(\mathcal{G})\right]}
    \label{eq:a}
\end{equation}
    \label{thm:canonical_ordering}
\end{theorem}

\paragraph{Proof of Theorem \ref{thm:canonical_ordering}.} 
We start with implication (Eq.~\ref{eq:canonical_implication_1}), as it is easier to show.
Assume that $\mathcal{G}$ and $\mathcal{G}'$ are isomorphic under $\mathcal{H}$.
Two graphs $(\mathcal{G}, \mathcal{G}')$ can be isomorphic only if they have the same number of nodes. Let $n = |\mathcal{V}| =|\mathcal{V}'|$.
We first show that, in-between and after calculating equations \ref{eq:H_j} then \ref{eq:H_j_update} then \ref{eq:H_j_update2},  the following \textbf{property} is maintained: matrices $\mathbf{H}^\mathcal{H}$ and ${\mathbf{H}'}^\mathcal{H}$
contain the same rows, but not necessarily in the same order. Then, we show that left-multiplication with $\mathbf{A}$ sorts rows with matching orders.
\begin{itemize}[topsep=0pt]
    \item Since $(\mathcal{G}, \mathcal{G}')$ are assumed isomorphic under $\mathcal{H}$, therefore $\mathcal{X}$ is just a re-ordering of $\mathcal{X}'$ (per Definition \ref{definition:isomorphism_feats}.
    Since $ \mathbf{H}_j =  \textrm{\$}( \mathcal{X}_j)$ and $ \mathbf{H}'_j =  \textrm{\$}( \mathcal{X}'_j)$, then $\mathbf{H}$ is just a re-ordering of $\mathbf{H}'$ and therefore the property is maintained after 
    Eq.~\ref{eq:H_j}.
    \item To prove the property is maintained after calculating Eq.~\ref{eq:H_j_update} follows.
    \textsc{TopologicalOrder} processes every node exactly once.
    Starting from nodes $j$ where $|\mathcal{P}_j| = 0$,
    the update $\mathbf{H}_j^\mathcal{H} := \$ \left(
    \mathbf{H}_j^\mathcal{H} \oplus \texttt{sort}( \{ \mathbf{H}_k^\mathcal{H} \mid (k, j) \in \mathcal{E} \})   \right)$ reduces to $\mathbf{H}_j^\mathcal{H} := \$ \left(
    \mathbf{H}_j^\mathcal{H}\right)$.
    More generally, after computing Eq.\ref{eq:H_j_update} for any $j$, \textsc{TopologicalOrder} guarantees that the row $\mathbf{H}^\mathcal{H}_j$
    is exactly a function of $\mathcal{P}_j$ (when restricting to features in $\mathcal{H}$).
    \item The proof that property is maintained after calculating Eq.~\ref{eq:H_j_update2} mirrors the above but following reverse-topological order of $\mathcal{S}$ in lieu of $\mathcal{P}$.
    
    % Afterwards, when $j$ is about to be processed, it must be that all its incoming neighbors $\{u \mid (u, j) \in \mathcal{E}\}  \subseteq \mathcal{P}_j$ and therefore they have already been processed in previous iterations.
    % %
    
    % Therefore, for every $j \in \mathcal{V}$ where $|\mathcal{P}_j|=0$, the node hash $\mathbf{H}_j$ is only a function of the node. Since graphs are isomorphic (under feature set $\mathcal{H}$), there must be the same number of leaf nodes with the same features (under $\mathcal{H}$) and therefore the same entries must appear in ${\mathbf{H}'}^\mathcal{H}$. 

    % \item We prove, by induction, that the property is maintained after executing Eq.~\ref{eq:H_j_update}. The induction follows the topological order.
    % \underline{Basis Step.} \textsc{TopologicalOrder} starts by choosing nodes $j$ where $|\mathcal{P}_j| = 0$.
    % The update $\mathbf{H}_j^\mathcal{H} := \$ \left(
    % \mathbf{H}_j^\mathcal{H} \oplus \texttt{sort}( \{ \mathbf{H}_k^\mathcal{H} \mid (k, j) \in \mathcal{E} \})   \right)$ reduces to $\mathbf{H}_j^\mathcal{H} := \$ \left(
    % \mathbf{H}_j^\mathcal{H}\right)$. Therefore, for every $j \in \mathcal{V}$ where $|\mathcal{P}_j|=0$, the node hash $\mathbf{H}_j$ is only a function of the node. Since graphs are isomorphic (under feature set $\mathcal{H}$), there must be the same number of leaf nodes with the same features (under $\mathcal{H}$) and therefore the same entries must appear in ${\mathbf{H}'}^\mathcal{H}$. \underline{Induction Hypothesis.} When an entry 
\end{itemize}
Finally, the multiplication $\mathbf{A} \times \mathbf{H}$ only re-orders the nodes of $\mathbf{H}$ (per Eq.~\ref{eq:a}), exactly to sort the rows of $\mathbf{H}$ lexicographically (per Eq.~\ref{eq:h_and_pi}). This applies to both $\mathbf{H}^\mathcal{H}(\mathcal{G})$ and $\mathbf{H}^\mathcal{H}(\mathcal{G}')$.
\begin{equation*}
\textrm{Therefore,} \ \ \ \ \ \ \ \ \ \ \ 
\mathcal{G} \underset{\mathcal{H}}{\cong} \mathcal{G}' \ \  \Longrightarrow \ \
    \mathbf{A}^{\pi^\mathcal{H}(\mathcal{G})} \times  \mathbf{H}^\mathcal{H}(\mathcal{G})  = 
    \mathbf{A}^{\pi^\mathcal{H}(\mathcal{G}')} \times  \mathbf{H}^\mathcal{H}(\mathcal{G}').
\end{equation*}

We prove the reverse implication (Eq.~\ref{eq:canonical_implication_2}) by contradiction.
\begin{align}
\textrm{For the sake of contradiction, assume: }\ \ \ \  &
    \mathbf{A}^{\pi^\mathcal{H}(\mathcal{G})} \times  \mathbf{H}^\mathcal{H}(\mathcal{G})  = 
    \mathbf{A}^{\pi^\mathcal{H}(\mathcal{G}')} \times  \mathbf{H}^\mathcal{H}(\mathcal{G}'), \label{eq:thm1contradictiveassumption} \\
\textrm{and not: }\ \ \ \ &
\mathcal{G} \underset{\mathcal{H}}{\cong} \mathcal{G}'.
\end{align}

The assumption (Eq.~\ref{eq:thm1contradictiveassumption}) implies that every for any row $j\in \mathcal{V}$, the string (bit vector) $\mathbf{H}_j^{\mathcal{H}}(\mathcal{G}) \in \{0, 1\}^{256}$   exists at some row in $\mathbf{H}^{\mathcal{H}}(\mathcal{G}')$.
We now show that $\mathbf{H}_j^{\mathcal{H}}(\mathcal{G})$ is a deterministic uniform-random function of
$\{\mathcal{X}_k[(t, a)] \ \  \mid \ \  k\in \{j\} \cup \mathcal{P}_i \cup \mathcal{S}_i  \}_{(t,a) \in \mathcal{H} }$, plus the edge structure of $\{j\} \cup \mathcal{P}_i \cup \mathcal{S}_i$ that is linking these feature nodes.
Crucially, a bijective function, with high probability (\textit{whp}).

When calculating $\mathbf{H}^\mathcal{H}(G)$,
each row $\mathbf{H}^\mathcal{H}_j$ will be updated once in each of 
Equations \ref{eq:H_j}, \ref{eq:H_j_update}, and \ref{eq:H_j_update2}, \textit{i.e.}, thrice.
First updates (Eq.\ref{eq:H_j}) can happen to all nodes in-parallel. Second updates (Eq.\ref{eq:H_j_update}) happen in topological order, and third updates happen in reverse-topological order (Eq.\ref{eq:H_j_update2}).
\begin{itemize}[topsep=0pt]
    \item After first set of updates  (Eq.~\ref{eq:H_j}),  $ \mathbf{H}_j^\mathcal{H} = \$\left( \oplus\{\mathcal{X}_j[(t, a)]  \}_{(t, a) \in \mathcal{H}} \right)$ encorporate into $\mathbf{H}_j$ the features of nodes $\{j\}$.
    \item The second set of updates proceeds in topological order. For leaf nodes, they will just re-hash their their features \textit{i.e.} $\mathbf{H}_j = \$\left(\$\left( \{\mathcal{X}_j[(t, a)] \}\right) \right) $. Subsequent (non-leaf node) node $j$ updates its hash, by concatenating the current $\mathbf{H}_j$ (already capturing $\mathcal{X}_j$),
    with already updated hashes of their incoming neighbors $\{\mathbf{H}_k\}_{(k, j) \in \mathcal{E}}$. This update includes the in-degree \textit{local structure}. Since each neighbor $\mathbf{H}_k$ has already updated from its predecessor neighbors, then recursively and by induction, each node $j$ updates its hash to a deterministic function of features of all nodes  $\in \{j\}\cup\mathcal{P}_j$.
    \item Echoing the above, but in reverse topological order, updates 
    string $\mathbf{H}_i$ to 
    its final value, a deterministic function of features of nodes all nodes  $\in \{j\}\cup\mathcal{P}_j\cup\mathcal{S}_j$.
\end{itemize}
It is important to realize that hashing function $\$(.)$ is run on its own output (like $\$( \$(.))$. We wish to have the output to be uniform -- \textit{i.e.}, each outcome has $\approx \frac{1}{2^{256}}$ to appear. We  are therefore restricted to cryptographic hashing functions. In practice, we use MD5. This shows that:
\begin{equation}
    \mathbf{A}^{\pi^\mathcal{H}(\mathcal{G})} \times  \mathbf{H}^\mathcal{H}(\mathcal{G})  = 
    \mathbf{A}^{\pi^\mathcal{H}(\mathcal{G}')} \times  \mathbf{H}^\mathcal{H}(\mathcal{G}') \ \  
    \underset{whp}{\Longrightarrow} \ \
    \mathcal{G} \underset{\mathcal{H}}{\cong} \mathcal{G}' 
\end{equation}
\qed
%\hspace{1cm} \unskip\nobreak\hfill $\square$

\begin{theorem}
    The sets $\mathcal{H} \subseteq \mathcal{A}$ and $\mathcal{H} \subseteq \mathcal{A}$ can extract a canonical feature vector. Specifically,
\begin{equation}
\mathcal{G} \underset{(\mathcal{H} \cup \mathcal{F})}{\cong} \mathcal{G}'
\ \  \Longrightarrow \ \
\mathbf{x}^\mathcal{H}_\mathcal{F}(\mathcal{G}) = \mathbf{x}^\mathcal{H}_\mathcal{F}(\mathcal{G}')
\end{equation}
\label{thm:canonical_feature}
\end{theorem}

\paragraph{Proof of Theorem \ref{thm:canonical_feature}.} We copy Eq.~\ref{eq:x}:
\begin{equation*}
\mathbf{x}^{\mathcal{H}}_{\mathcal{F}} = \bigoplus_{j \in \pi^\mathcal{H}} \left\{ f_{(t, a)}(\mathcal{X}_j[(t, a)])   \ \ \mid \ \ t = \tau_j \right\}_{(t, a) \in \mathcal{F}}
%\label{eq:x}
\end{equation*}
which rasterizes node features into a flat vector, using the ordering dictated by $\pi^\mathcal{H}(G)$. We are given that: $\mathcal{G} \underset{(\mathcal{H} \cup \mathcal{F})}{\cong} \mathcal{G}'$. But,
\begin{equation*}
    \mathcal{G} \underset{(\mathcal{H} \cup \mathcal{F})}{\cong} \mathcal{G}'\ \ \Longrightarrow \ \  \mathcal{G} \underset{\mathcal{H}}{\cong} \mathcal{G}'
\end{equation*}
as the right-side is less restrictive.
Using Theorem\ref{thm:canonical_ordering},
$\pi^\mathcal{H}(\mathcal{G})$ corresponds to
$\pi^\mathcal{H}(\mathcal{G}')$,
specifically equating 
\begin{equation}
\bigotimes_{j \in \pi^\mathcal{H}(\mathcal{G})} \left\{ \psi(\mathcal{X}_j ) \right\} = \bigotimes_{j \in \pi^\mathcal{H}(\mathcal{G}')} \left\{ \psi(\mathcal{X}'_j ) \right\} 
\label{eq:thm_general_featurizer}
\end{equation}
for any arbitrary function $\psi(.)$ and any (ordered set) aggregation function $\otimes$.
Choosing $\otimes$ as  = $\oplus$ and $\psi(\ .) = \left\{ f_{(t, a)}(\ .[(t, a)])   \ \ \mid \ \ t = \tau_j \right\}_{(t, a) \in \mathcal{F}}$ recovers that
$
\mathbf{x}^{\mathcal{H}}_{\mathcal{F}}(\mathcal{G}) = 
\mathbf{x}^{\mathcal{H}}_{\mathcal{F}}(\mathcal{G}')
$. Therefore,
\begin{equation*}
\mathcal{G} \underset{(\mathcal{H} \cup \mathcal{F})}{\cong} \mathcal{G}'
\ \  \Longrightarrow \ \
\mathbf{x}^\mathcal{H}_\mathcal{F}(\mathcal{G}) = \mathbf{x}^\mathcal{H}_\mathcal{F}(\mathcal{G}')
\end{equation*}
\qed

\begin{theorem}
    Given an arbitrary anchor graph $\mathcal{G}$, then every $\mathbf{x} \in \{ \mathbf{x}^\mathcal{H}_\mathcal{F}(\mathcal{G}') \mid h(\mathcal{G}) = h(\mathcal{G}') \}$ has the same dimensionality, with canonical node-to-feature positions.
    \label{thm:cross_learning}
\end{theorem}

\paragraph{Proof of Theorem \ref{thm:cross_learning}}
From Theorem \ref{thm:canonical_ordering}, we have:
\begin{equation*}
    \mathbf{A}^{\pi^\mathcal{H}(\mathcal{G})} \times  \mathbf{H}^\mathcal{H}(\mathcal{G})  = 
    \mathbf{A}^{\pi^\mathcal{H}(\mathcal{G}')} \times  \mathbf{H}^\mathcal{H}(\mathcal{G}')
    \ \ 
    \underset{whp}{\Longrightarrow} 
    \ \ 
    \mathcal{G} \underset{\mathcal{H}}{\cong} \mathcal{G}'
\end{equation*}
    
Moreover, we have that:
\begin{equation}
    \mathbf{A}^{\pi^\mathcal{H}(\mathcal{G})} \times  \mathbf{H}^\mathcal{H}(\mathcal{G})  = 
    \mathbf{A}^{\pi^\mathcal{H}(\mathcal{G}')} \times  \mathbf{H}^\mathcal{H}(\mathcal{G}')
    \ \ 
    \Longrightarrow
    \ \ 
    h^\mathcal{H} (\mathcal{G}) = h^\mathcal{H} (\mathcal{G}'),
    \label{eq:thm_nodelevel_h_implies_graphlevel_h}
\end{equation}
which follows from the definition of $h^\mathcal{H}(.)$ in Eq.~\ref{eq:h_and_pi} as:
\begin{align*}
    h^\mathcal{H} (\mathcal{G}) = \textrm{\Large \$}\left( \bigoplus_{j \in \pi^{\mathcal{H}}} \mathbf{H}_j^\mathcal{H}(\mathcal{G}) \right) &= \textrm{\Large \$} \left( \bigoplus_{j \in \{1, 2, \dots, n\}}  \left[ \mathbf{A}^{\pi^\mathcal{H}(\mathcal{G})} \times  \mathbf{H}^\mathcal{H}(\mathcal{G}) \right]_j \right) \\
    & = \textrm{\Large \$} \left( \bigoplus_{j \in \{1, 2, \dots, n\}}  \left[ \mathbf{A}^{\pi^\mathcal{H}(\mathcal{G}')} \times  \mathbf{H}^\mathcal{H}(\mathcal{G}') \right]_j \right) = h^\mathcal{H} (\mathcal{G}')
\end{align*}

The converse of Eq.~\ref{eq:thm_nodelevel_h_implies_graphlevel_h} holds with high probability, specifically, since \$ is a uniform hashing function, \textit{i.e.}, producing 1-to-1 mapping (with collision rate of $\frac{1}{2^{256}}$). Therefore, we have:
\begin{align*}
    h^\mathcal{H} (\mathcal{G}) = h^\mathcal{H} (\mathcal{G}')
    \ \ 
    &\underset{whp}{\Longrightarrow} 
    \ \ 
    \mathbf{A}^{\pi^\mathcal{H}(\mathcal{G})} \times  \mathbf{H}^\mathcal{H}(\mathcal{G})  = 
    \mathbf{A}^{\pi^\mathcal{H}(\mathcal{G}')} \times  \mathbf{H}^\mathcal{H}(\mathcal{G}')
    \\
    \textrm{hence,  } \ \ \ 
    h^\mathcal{H} (\mathcal{G}) = h^\mathcal{H} (\mathcal{G}')
    \ \ 
    &\underset{whp}{\Longrightarrow} 
    \ \ 
    \mathcal{G} \underset{\mathcal{H}}{\cong} \mathcal{G}'.
\end{align*}

%The converse also holds \textit{whp}
%since $h^\mathcal{H} = \textrm{\Large \$}\left( \bigoplus_{j \in \pi^{\mathcal{H}}} \mathbf{H}_j^\mathcal{H} \right)$

% \begin{equation}
%     h^\mathcal{H} = \textrm{\Large \$}\left( \bigoplus_{j \in \pi^{\mathcal{H}}} \mathbf{H}_j^\mathcal{H} \right),  \ \ \ \  \ \ \ \textrm{with} \ \ \ \ \ \pi^{\mathcal{H}} = \arg\textrm{sort}(\{ \mathbf{H}_j^\mathcal{H} \}_{j \in \mathcal{V}}).
% \label{eq:h_and_pi}
% \end{equation}

Finally,  Theorem \ref{thm:cross_learning} considers pairs for which $h(\mathcal{G}) = h(\mathcal{G}')$. Therefore, with high probability (due to above), $\mathcal{G} \underset{\mathcal{H}}{\cong} \mathcal{G}$.
Therefore, the ordering $\pi^\mathcal{H}(\mathcal{G})$ must be consistent with $\pi^\mathcal{H}(\mathcal{G}')$.
The sequence of node \textbf{types}, when iterating over $\mathcal{G}$  per $\pi^\mathcal{H}(\mathcal{G})$, must be the same sequence of node types when iterating over $\mathcal{G}'$  per $\pi^\mathcal{H}(\mathcal{G}')$. During these iterations, the vectors $\mathbf{x}^\mathcal{H}_\mathcal{F}(\mathcal{G})$ and $\mathbf{x}^\mathcal{H}_\mathcal{F}(\mathcal{G}')$ are composed.
Since the feature dimension is deterministic given a node type, then (each type, structural position) will occupy distinct positions in the feature vectors.
\qed

As an aside, in our implementation, we also always include these features for all nodes: in-degree, out-degree, and node type (table, column, operand, ...) and always include them in $\mathcal{H}$.

\section{Feature Extractors}
\label{sec:appendix_feature_extractors}
We define several functions. Each can extract node features. For any node, its entire feature vector is the concatenation of all applicable feature extractors. We implement a handful of $f$'s:
\begin{enumerate}[label=($f_{\arabic*}$)]
\item $f_{\textrm{num}}(m) = m \in \mathbb{R}^1$. Applies to numeric literals. Casting from string to number is implied.\label{enum:fnum}

\item $f_{\textrm{scaled}}(m) = \frac{m - \texttt{minVal}(\mathbf{c})}{\texttt{maxVal}(\mathbf{c}) - \texttt{minVal}(\mathbf{c})} \in \mathbb{R}^1$. Applies to numeric literals when used alongside column $\mathbf{c}$. It can be activated if the DB engine stores min- and max-value per column.

\item $f_\textrm{comp}(m) \in \mathbb{R}^2 $ applies when literal is ordinally-compared with column $\mathbf{c}$ (with op $=, >, \ge, <, \le$). If op is $<$ or $\le$ then $f_\textrm{comp}(m) = [0, f_\textrm{scaled}(m)]$. If op is $>$ or $\ge$, then  $f_\textrm{comp}(m) = [f_\textrm{scaled}(m), 1]$. Finally, if op is $=$, then  $f_\textrm{comp}(m) = [f_\textrm{scaled}(m), f_\textrm{scaled}(m)].$

\item $f_\textrm{ASCII}(s) = \texttt{[ord(s[0]) ord(s[1]), ord(s[2])]} \in \mathbb{R}^3$. Applies to string literals, where \texttt{ord(.)} is the ASCII code of character \texttt{s[.]}.

\item $f_\textrm{date}(d) = [\texttt{d.year}, \texttt{d.month}, \texttt{d.day}] \in \mathbb{R}^3$. Applies to date literals.\label{enum:fdate}

\item $f_\textrm{tableSize}(\texttt{table}) = \texttt{table.size} \in \mathbb{R}^1 $. Applies for table nodes.

\item $f_\textrm{columnRange}(\mathbf{c})  = [\mathbf{c}.\texttt{minVal}, \mathbf{c}.\texttt{maxVal} ] \in \mathbb{R}^2 $. Applies for column nodes.

\item $f_\textrm{ordinalOp}(op) \in \{0, 1\}^3$. Applies to ordinal operations $=, >, \ge, <, \le$, respectively as $[010]$, $[001]$, $[011]$, $[100]$, $[110]$.
\end{enumerate}
We leave the design of more intricate $f$'s as future work. The \textbf{learning features}
\begin{equation}
\mathcal{F} \subset \{ (t, a, f) \ \ \mid \ \  (t, a) \in \mathcal{A}, \ \  f \in (\{0, 1\}^{*} \rightarrow \mathbb{R}^{*} ) \},
\end{equation}
allow us to customize how to extract numeric features from attribute $a$ node type $t \in \mathcal{T}$.

\section{Experiments, Ablation Studies, Discussions}
\label{appendix:experiments}
For ablation studies, we run experiments on CardBench workloads with increasing complexity, these datasets are downloaded from benchmark~\cite{cardbench}.
% Further, we extend their query generator to: 1) enable  multi-way join queries (up-to 5 joins) to increase the query complexity; 2) incorporate the high repetiveness feature of data warehouse workloads as in Redshift~\cite{redshift} (prefixed ``multijoin-''). 

%For all multijoin datasets, we fixed the sample constant size at 10 and varied the sample size (repetition rate) to evaluate its impact on accuracy in Fig ~\ref{fig:ablation_repetition}.

\begin{table}[b]
\centering
\caption{Q-Error Comparison on CardBench Workloads.}
\label{tab:qerror_multijoin}
\footnotesize
\begin{tabular}{l c c c c c c}
\toprule
Model & \multicolumn{3}{c}{cms} & \multicolumn{3}{c}{stackoverflow} \\
\cmidrule(lr){2-4} \cmidrule(lr){5-7}
& $Q_\textrm{err}^{50}$ & $Q_\textrm{err}^{90}$ & $Q_\textrm{err}^{95}$ & $Q_\textrm{err}^{50}$ & $Q_\textrm{err}^{90}$ & $Q_\textrm{err}^{95}$ \\
\midrule
Postgres & $3.33$ & $112$ & $2.3e^{3}$ & $4.85$ & $360$ & $3.1e^{3}$ \\
$(H_3, \texttt{P})$ & $3.21$ & $110$ & $2.2e^{3}$ & $4.30$ & $367$ & $3.8e^{3}$ \\
$(H_2, H_3, \texttt{P})$ & $1.15$ & $46.67$ & $159$ & $1.16$ & $44.33$ & $464$ \\
$(H_1, \texttt{P})$ & $1.07$ & $22.22$ & $97.00$ & $1.12$ & $21.03$ & $200$ \\
$(H_1, H_2, \texttt{P})$ & $\mathbf{1.06}$ & $\mathbf{20.10}$ & $\mathbf{94.48}$ & $\mathbf{1.11}$ & $\mathbf{18.01}$ & $\mathbf{182}$ \\
$(H_1, H_2, H_3, \texttt{P})$ & $\mathbf{1.06}$ & $\mathbf{20.10}$ & $\mathbf{94.48}$ & $\mathbf{1.11}$ & $\mathbf{18.01}$ & $\mathbf{182}$ \\
\midrule
Model & \multicolumn{3}{c}{accidents} & \multicolumn{3}{c}{airline} \\
\cmidrule(lr){2-4} \cmidrule(lr){5-7}
& $Q_\textrm{err}^{50}$ & $Q_\textrm{err}^{90}$ & $Q_\textrm{err}^{95}$ & $Q_\textrm{err}^{50}$ & $Q_\textrm{err}^{90}$ & $Q_\textrm{err}^{95}$ \\
\midrule
Postgres & $1.65$ & $10.31$ & $18.29$ & $1.63$ & $97.30$ & $216$ \\
$(H_3, \texttt{P})$ & $1.34$ & $8.93$ & $20.60$ & $1.59$ & $97.00$ & $216$ \\
$(H_2, H_3, \texttt{P})$ & $1.15$ & $\mathbf{4.81}$ & $\mathbf{15.42}$ & $1.20$ & $13.88$ & $91.00$ \\
$(H_1, \texttt{P})$ & $1.15$ & $4.95$ & $17.25$ & $1.13$ & $4.50$ & $29.20$ \\
$(H_1, H_2, \texttt{P})$ & $\mathbf{1.15}$ & $5.02$ & $17.70$ & $\mathbf{1.13}$ & $\mathbf{4.29}$ & $\mathbf{25.00}$ \\
$(H_1, H_2, H_3, \texttt{P})$ & $\mathbf{1.15}$ & $5.02$ & $17.70$ & $\mathbf{1.13}$ & $\mathbf{4.29}$ & $\mathbf{25.00}$ \\
\midrule
Model & \multicolumn{3}{c}{employee} & \multicolumn{3}{c}{geo} \\
\cmidrule(lr){2-4} \cmidrule(lr){5-7}
& $Q_\textrm{err}^{50}$ & $Q_\textrm{err}^{90}$ & $Q_\textrm{err}^{95}$ & $Q_\textrm{err}^{50}$ & $Q_\textrm{err}^{90}$ & $Q_\textrm{err}^{95}$ \\
\midrule
Postgres & $1.54$ & $3.38$ & $4.83$ & $224$ & $2.1e^{5}$ & $1.2e^{6}$ \\
$(H_3, \texttt{P})$ & $1.35$ & $3.14$ & $4.42$ & $218$ & $2.1e^{5}$ & $1.2e^{6}$ \\
$(H_2, H_3, \texttt{P})$ & $1.05$ & $2.11$ & $\mathbf{2.98}$ & $1.10$ & $5.8e^{3}$ & $7.3e^{4}$ \\
$(H_1, \texttt{P})$ & $1.03$ & $2.09$ & $3.07$ & $1.09$ & $192$ & $1.1e^{4}$ \\
$(H_1, H_2, \texttt{P})$ & $\mathbf{1.03}$ & $\mathbf{2.03}$ & $3.01$ & $\mathbf{1.08}$ & $\mathbf{66.38}$ & $\mathbf{7.0e^{3}}$ \\
$(H_1, H_2, H_3, \texttt{P})$ & $\mathbf{1.03}$ & $\mathbf{2.03}$ & $3.01$ & $\mathbf{1.08}$ & $\mathbf{66.38}$ & $\mathbf{7.0e^{3}}$ \\
\bottomrule
\end{tabular}
\end{table}

\subsection{Hierarchical Models} 
We first examine the effectiveness and necessity of keeping multiple hierarchies in \ours. 
Table ~\ref{tab:qerror_multijoin} compares the Q-Error metrics of different hierarchy configurations (using various combinations of $\mathcal{H}_1, \mathcal{H}_2, \mathcal{H}_3$) against \postgres on several CardBench datasets. 
% In this set of experiments, methods must always make a prediction.
% Our method defaults to the Postgres estimator, in cases, where the graph structure is novel (has not appeared earlier in the online setting).
%Our full hierarchy, depicted in Figure~\ref{fig:template_hierarchy} and formalized in Equation~\ref{eq:mastermodel}, is abbreviated $(H_1, H_2, H_3, \texttt{P})$, where \texttt{P} denoting Postgres estimator. We set thresholds $(\tau_1, \tau_2, \tau_3)$ in Eq.\ref{eq:mastermodel} to (3, 10, 100) and employ Gradient-Boosted Decision Trees (GBDT) at each hierarchical level. Table~\ref{tab:qerror_multijoin} compares hierarchical models with different hierarchy combinations.
The table shows that progressively incorporating more granular hierarchy levels ($\mathcal{H}_3$, $\mathcal{H}_2$, then $\mathcal{H}_1$) consistently improves estimation accuracy across datasets and percentiles. For instance, on `cms' workload, the P90 Q-error improves from 112 (Postgres) to 110 $(\mathcal{H}_3, \texttt{P})$, then to 46.67 $(\mathcal{H}_2, \mathcal{H}_3, \texttt{P})$, and finally to 20.10 $(\mathcal{H}_1, \mathcal{H}_2, \texttt{P})$ or $(\mathcal{H}_1, \mathcal{H}_2, \mathcal{H}_3, \texttt{P})$.
% Comparing $(H_1, H_2, H_3, \texttt{P})$,  $(H_2, H_3, \texttt{P})$,  $(H_3, \texttt{P})$, and Postgres, we can see the models keep improving when we add more levels of hierarchy and the full hierarchy of models is always better than Postgres at all metrics.
%In addition, The full hierarchy leverages each level effectively, as evidenced by the activation ratios (0.69, 0.04, 0.01, 0.26) for $H_1$, $H_2$, $H_3$, and Postgres, respectively.
These results demonstrate the effectiveness of our hierarchical models in leveraging historical data to enhance the cardinality estimation capabilities of traditional optimizers. Moreover, Table ~\ref{tab:qerror_multijoin} shows the need for multiple hierarchies. Comparing $(H_1, \texttt{P})$, $(\mathcal{H}_1, \mathcal{H}_2, \texttt{P})$, $(\mathcal{H}_1, \mathcal{H}_2, \mathcal{H}_3, \texttt{P})$, the latter two consistently outperform the first. This indicates that a simple hierarchy $(\mathcal{H}_1, \texttt{P})$ is insufficient, highlighting the importance of multi-level hierarchies.

\subsection{Model Choice}

\begin{figure}[b]
\vspace{-5pt}
    \centering
    % plot generated with python3 experimental/sysres/data_analytics/history_based/combine_results.py
    \includegraphics[width=0.7\linewidth]{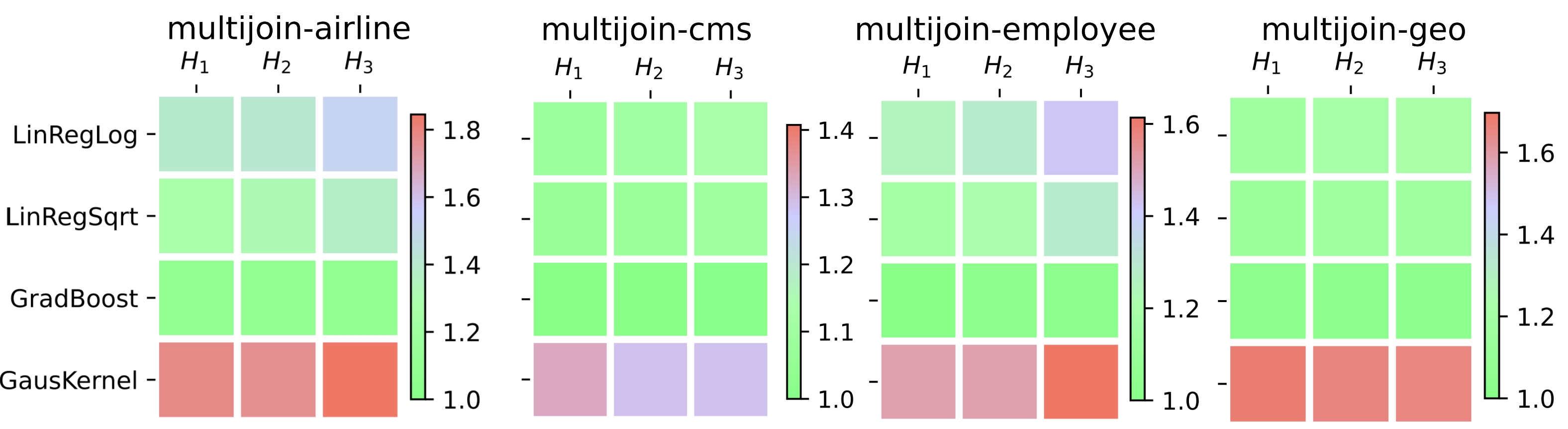}
    \caption{P50 Q-Error per database, comparing templatization strategies and learners.}
    \label{fig:heatmap}
    \vspace{-5pt}
\end{figure}

Figure~\ref{fig:heatmap} presents 50th percentile Q-errors comparing learned models (Linear Regression variants, Gradient Boosting, Gaussian Kernel) across hierarchy levels and datasets. Lower Q-errors are greener. The heatmap shows Gradient-Boosted Decision Trees (GBDT) achieve lowest median Q-errors, indicating superior accuracy.
%In terms of E2E time, while a Linear Regression model results in 50382.33 s on the 5000-query IMDb workload,
GBDT's E2E time is 49895s in Table ~\ref{tab:performance}, adding an overhead much smaller than savings due to better-optimized plans. 
Combined with efficient inference, GBDT was selected as the primary learner for \ours's overall evaluation (Table~\ref{tab:performance}, Table~\ref{tab:qerror_multijoin}).

% \subsubsection{Repetition Rate}
% We modify the workload generator in ~\cite{cardbench} to enable more constants for each predicate in the query. For example, instead of generating a query with predicates ``a $>$ 5 AND b $=$ 2", our modified generator will generate ``a $>$ 5 AND b $=$ 2", ``a $>$ 5 AND b $=$ 20", ``a $>$ 1 AND b $=$ 2", ``a $>$ 1 AND b $=$ 20" when the sample size is 2, meaning that each predicate will have 2 constants to choose from (ie. a $>$ [1, 10], b $=$ [2, 20]).
% The constant sample sizes in the experiment we choose are [1, 3, 10], therefore it generates the repetition rate of 20\%, 81\% and 91\% in query templates.
% As shown in Figure~\ref{fig:ablation_repetition}, all templatization strategies exhibit improved performance with increasing workload repetition, while maintaining low q-error levels.

% \begin{figure}[H]
%     \centering
%     \vspace{-0.5cm}
%     \includegraphics[width=\linewidth]{fig/sweep_history.pdf}
%     \caption{Accuracy of our learners, as a function of repetition amount. Each chart shows one templatization strategy, containing 4 lines: \{Gradient Boosted Decision Tree (Eq.~\ref{eq:xgboost_model}), Postgres Estimator \} $\times$ \{$50^\textrm{th}$, $90^\textrm{th}$ Q-errors\}. The Y-axis displays Q-errors.}
%     \label{fig:ablation_repetition}
%     \vspace{-0.5cm}
% \end{figure}

\subsection{History Size}

Figure ~\ref{fig:history_size} shows the impact of accumulated history size on \ours's estimation accuracy (P50 and P90 Q-Errors) on the IMDb workload. History size is less than or equal to x-axis value. The figure clearly shows that both P50 and P90 Q-Errors decrease significantly as the history size increases, especially in the initial stages. For instance, the P90 Q-Error drops sharply from over 200 towards 100 as history accumulates. The error curves then flatten, indicating that accuracy stabilizes once sufficient data is gathered for a template. This directly validates that \ours's learned models become more accurate as they are exposed to more examples through online learning.

% We also conduct ablation experiments to show that, in general, our simple models improve as data accumulates in each template (Figure~\ref{fig:q_error_vs_bucket_size}). As $H_1$ is the most-grained, it stabilizes earlier and has lower tail errors. Notably, the accuracy of coarser templatization, \textit{e.g.}, $H_3$, combining records from multiple (columns, predicate operators), needs more training history data to converge. 
% It also shows that GBDT always has better performance than Linear Regression (LR) and Gaussian Kernel(GK) models accross different datasets. This also matches our observation in Figure~\ref{fig:heatmap}.
% %We suspect it is a sign of under-fitting. This suggests that at the coarser nodes, one could train more complicated models, like neural methods (at some cadence), \textcolor{blue}{left as future work}.
% \begin{figure}[t]
%     \centering
%     \includegraphics[width=1\linewidth]{fig/qerror_upto_bucket_size.pdf}
%     \caption{Each subplot shows Q-error percentiles as function of amount of history per workload \& templatization strategy.
%     In particular,
%     each line color represents learner (Eq.\ref{eq:lr_model}--\ref{eq:xgboost_model}) and each line style represents percentile. History size is less than or equal to x-axis value.}
%     \label{fig:q_error_vs_bucket_size}
% \end{figure}

\subsection{Estimator Reliance Shift with Accumulated History}

%\begin{figure}[t]
%  \centering
 %   \includegraphics[width=0.4\linewidth]{fig/proportion.pdf}
  %  \caption{Shows the proportion of methods \ours used to make a estimate on the 5k IMDb workload. As we collect more queries, we rely less on \postgres estimates.}
  %  \label{fig:proportion}
%\end{figure}

Figure~\ref{fig:proportion} shows the proportion of subquery estimates from learned models vs. base \postgres as cumulative processed queries (history) increase on the 5k IMDb workload. The figure clearly demonstrates reliance shifting from \postgres (decreasing proportion) towards learned models (increasing proportion) as more history is gathered. This confirms \ours's online learning effectively leverages history to replace base estimates, underpinning iterative performance gains (Figure~\ref{fig:multi_iteration}).

\section{Runtime Analysis}
\sparagraph{Minimal Training Overhead Enables Online Learning.}
Table ~\ref{tab:performance} and Figure ~\ref{fig:scatter}  presents the total training overheads for all learned techniques. Offline, batch-trained methods like \mscn, \deepdb, \factorjoin, and fine-tuned \price incur substantial overheads, ranging from 1,466 seconds (\mscn) to 14,828 seconds (\price fine-tuned). Note these exclude data collection costs for query-driven methods $\approx 34$ hours for \mscn). Such high costs impede frequent updates. In contrast, \ours, an online learner, starts with zero initial overhead and incurs a total training overhead of only 37.29s for the 5k workload via lightweight incremental updates ($\approx 0.001$s each). These updates can be performed asynchronously.

This minimal overhead enables practical online learning and continuous adaptation, fundamentally distinguishing \ours from expensive batch retraining paradigms.

\subsection{Detailed Analysis}
\label{sec:expr_detail}

\sparagraph{Detailed Runtime Comparison.}
Figure~\ref{fig:detailed_runtime} shows the relative End-to-End time improvement over \postgres (0\% line) for queries grouped by their original PG runtime. For very short queries ([0-0.008s], [0.008-0.66s]), most learned methods show degradation, as optimization time dominates. PRICE exhibits the largest degradation, while \ours stays close to \postgres and even shows a slight initial improvement. For longer queries (especially $>$200s), where execution time is substantial, learned methods like \deepdb, \factorjoin, and \ours achieve significant improvements, as the benefit of better estimates outweighs optimization overhead.
This demonstrates that low optimization overhead is crucial for performance on short queries, while estimation accuracy drives improvements on long ones.
Figure~\ref{fig:detailed_runtime} confirms \ours provides robust performance across query runtimes, avoiding degradation on short queries due to its low optimization cost, while delivering substantial gains on long queries.

\sparagraph{Relative Estimation Error Distribution.}
Figure~\ref{fig:q-error} shows the distribution of relative estimation errors (estimated/true) for all 46,928 subqueries on the 5000-query IMDb workload. Perfect estimates are at 1. The figure reveals \postgres and \price estimates are heavily skewed below 1, indicating significant underestimation bias. In contrast, \ours, \deepdb, \factorjoin, and \mscn distributions are centered around 1, showing reduced bias. \ours and \deepdb exhibit the tightest distributions around 1, signifying lower error variance. Such reduced bias and variance are crucial for effective query optimization.
Figure~\ref{fig:q-error} demonstrates \ours significantly improves estimation accuracy and reduces the underestimation bias compared to PostgreSQL.

% We also evaluated the impact of this bias adjustment on a larger workload. On the first 1000 queries of the 5000-query IMDb workload, the End-to-End time with unadjusted PostgreSQL bias was 17884.00 seconds. Applying our bias adjustment strategy reduced this to 11063.61 seconds, significantly improving performance and surpassing PostgreSQL's default of 13118.04 seconds for this workload subset. This large-scale result underscores the necessity of addressing PostgreSQL's inherent bias for effective learned estimator integration. The underestimate data required for this adjustment, as shown in Table~\ref{tab:pg_biased}, can be easily collected from executed queries with minimal or no extra overhead, making this a practical technique.

\sparagraph{Iterative Improvement through Online Learning.}
Figure~\ref{fig:multi_iteration} shows \ours's End-to-End time over 5 iterations on the first 1000 IMDb queries, compared to static baselines. \ours demonstrates a clear performance improvement trend, decreasing from $\approx 11,200$ seconds at Iteration 1 to $\approx 9,500$ seconds by Iteration 5. It starts faster than \postgres and \mscn, matches \factorjoin and \price early, and approaches \deepdb and \true performance over time. This improvement stems from effective online learning, where \ours refines its models with each processed query.
Figure~\ref{fig:multi_iteration} demonstrates that \ours's online learning delivers iterative End-to-End performance improvements, allowing it to adapt and become increasingly competitive with static learned estimators.

%Figure ~\ref{fig:end2end} shows our end-to-end performance on the IMDb dataset. Notably, if we apply our models directly on the PG default model results in suboptimal end-to-end performance. Iteration 0 took 17522.95 seconds which is longer than \postgres. This performance degradation arises from the mixing of \postgres and our model's estimates when estimating subquery cardinalities.
%~\textcolor{red}{Add an example to illustrate this?}
%\postgres numbers are often underestimating and thus tend to be chosen when \postgres plan the query, leading to long latency time. Table ~\ref{tab:pg_biased} illustrates the inherent bias in \postgres estimates. Notably, for queries involving 2 to 4 joins, the proportion of underestimated cardinalities exceeds 80\%. Moreover, as the number of joins increases, both the proportion of underestimated queries and the average Q-error increase. This is the known bias of underestimate in \postgres. ~\cite{?} To mitigate this bias, we introduced an adjustment to the PostgreSQL estimates by applying multiplicative factors, specific to the number of joins.

%Figure ~\ref{fig:end2end} shows that when mixing \ours and adjusted \postgres numbers, it decreased runtime to 11063 seconds. Table ~\ref{tab:job_light_performance} show our performance improvements over \postgres on the IMDb dataset. Notably, we can reduce the total end-to-end runtime from 13118 seconds to 9900 seconds at Iteration 1 (improves 25\% of the total runtime), while the true cardinality injection reduced from 13118 seconds to 8537 seconds (a 35\% performance boost).

\end{document}